\documentclass[12pt]{article}
\usepackage{amssymb,amsmath,amsthm,amsfonts,amscd}
 \usepackage{graphicx}
\newcommand\be{\begin{equation}}
\newcommand\ee{\end{equation}}
\newcommand\bea{\begin{eqnarray}}
\newcommand\eea{\end{eqnarray}}
\newcommand\ket[1]{|#1\rangle}

\newcommand{\fatalpha}{{\bf \alpha \kern -0.44em \alpha}}
\newcommand{\fatsigma}{{\bf \sigma \kern -0.54em \sigma}}
\newcommand{\tpchi}{{\bf \chi \kern -0.35em \chi}}
\newcommand{\llambda}{{\bf \lambda \kern -0.45em \lambda}}

\renewcommand{\theequation}{\arabic{equation}}
\renewcommand{\theequation}{\thesection-\arabic{equation}}
\bibliography{plain}
\title{\bf {Calculating effective resistances on underlying
networks of association schemes}}\vspace{20mm}
\author{ M. A. Jafarizadeh$^{a,b,c}$
\thanks{E-mail:jafarizadeh@tabrizu.ac.ir} ,
R. Sufiani$^{a,b}$
\thanks{E-mail:sofiani@tabrizu.ac.ir},
S. Jafarizadeh$^{d}$
\\ $^a${\small Department of Theoretical Physics and Astrophysics,
University of Tabriz, Tabriz 51664, Iran.} \\ $^b${\small
Institute for Studies in Theoretical Physics and Mathematics,
Tehran 19395-1795, Iran.} \\ $^c${\small Research Institute for
Fundamental Sciences, Tabriz 51664, Iran.} \\$^d${\small
Department of Electrical and computer engineering, University of
Tabriz, Tabriz 51664, Iran.}} \pagebreak

\vspace{20mm}
\begin{document}
\maketitle \vspace{15mm}
\newpage
\begin{abstract}
Recently, in Refs. \cite{jsj} and \cite{res2}, calculation of
effective resistances on distance-regular networks was investigated,
where in the first paper, the calculation was based on
stratification and Stieltjes function associated with the network,
whereas in the latter one a recursive formula for effective
resistances was given based on the Christoffel-Darboux identity. In
this paper, evaluation of effective resistances on more general
networks which are underlying networks of association schemes is
considered, where by using the algebraic combinatoric structures of
association schemes such as stratification and Bose-Mesner algebras,
an explicit formula for effective resistances on these networks is
given in terms of the parameters of corresponding association
schemes. Moreover, we show that for particular underlying networks
of association schemes with diameter $d$ such that the adjacency
matrix $A$ possesses $d+1$ distinct eigenvalues, all of the other
adjacency matrices $A_i$, $i\neq 0,1$ can be written as polynomials
of $A$, i.e., $A_i=P_i(A)$, where $P_i$ is not necessarily of degree
$i$. Then, we use this property for these particular networks and
assume that all of the conductances except for one of them, say
$c\equiv c_1=1$, are zero to give a procedure for evaluating
effective resistances on these networks. The preference of this
procedure is that one can evaluate effective resistances by using
the structure of their Bose-Mesner algebra without any need to know
the spectrum of the
adjacency matrices. 

{\bf PACs Index: 03.65.Ud }
\end{abstract}
\vspace{70mm}
\newpage
\section{Introduction}
A classic problem in electric circuit theory studied by numerous
authors over many years, is the computation of the resistance
between two nodes in a resistor network (see, e.g., \cite{Cserti}).
The effective resistance has a probabilistic interpretation based on
classical random walker walking on the network. Indeed, the
connection between random walks and electric networks has been
recognized for some time (see e.g. \cite{Kakutani, Kemeny, Kelly} ),
where one can establish a connection between the electrical concepts
of current and voltage and corresponding descriptive quantities of
random walks regarded as finite state Markov chains (for more
details see \cite{Doyle}). Also, by adapting the random-walk
dynamics and mean-field theory it has been studied that
\cite{Bosiljka}, how the growth of a conducting network, such as
electrical or electronic circuits, interferes with the current flow
through the underlying evolving graphs. In \cite{jss}, the authors
have been shown that, there is also connection between the
mathematical techniques for investigating CTQW on graphs, such as
Hilbert space of the walk based on stratification and spectral
analysis, and electrical concept of resistance between two arbitrary
nodes of regular networks and the same techniques can be employed
for calculating the resistance. Recently, in Refs. \cite{jsj} and
\cite{res2}, calculation of effective resistances on
distance-regular networks was investigated, where in the first
paper, the calculation was based on stratification and Stieltjes
function associated with the network, whereas in the latter one a
recursive formula for effective resistances was given based on the
Christoffel-Darboux identity for orthogonal polynomials. In this
paper, we consider more general resistor networks which are
underlying networks of association schemes. In fact, the theory of
association schemes \cite{Ass.sch.} (the term of association scheme
was first coined by R. C. Bose and T. Shimamoto in \cite{Bose}) has
its origin in the design of statistical experiments. The connection
of association schemes to algebraic codes, strongly regular graphs,
distance-regular graphs, design theory etc., further intensified
their study. A further step in the study of association schemes was
their algebraization. This formulation was done by R. C. Bose and D.
M. Mesner who introduced an algebra generated by the adjacency
matrices of the association scheme, known as Bose-Mesner algebra. We
will employ the algebraic structures of the underlying networks of
association schemes in order to calculate the effective resistances
between arbitrary nodes of them in terms of the parameters of the
corresponding association scheme such as diameter of the scheme, the
so-called first eigenvalue matrix $P$, the valencies of the
adjacency matrices and the rank of the corresponding idempotents. As
we will see, the preference of this employment is that we able to
give analytical formulas for effective resistances on these networks
in terms of the known parameters of the corresponding association
schemes. As it will be shown in section 5, in order to calculate the
effective resistances on underlying networks of association schemes,
one needs to know the spectrum of the adjacency matrices $A_i$ for
$i=1,...,d$. Although, in the most cases the spectrum of the
Bose-Mesner algebra is known (for example in the cases of group
association schemes), but the formulas for effective resistances in
terms of the spectrum of the networks do not possess a closed form
and evaluation of them in the most cases is not an easy task. So,
first we assume that all of the conductances except for one of them,
say $c\equiv c_1=1$, are zero and  consider particular underlying
networks of association schemes such that the adjacency matrices
$A_i$ can be written as polynomials of the first adjacency matrix
$A=A_1$ (not necessarily of degree $i$). Then, we give a procedure
for evaluating the effective resistances on these networks such that
one can calculate the effective resistances by using the structure
of their Bose-Mesner algebra without any need to know the spectrum
of the adjacency matrices.

The organization of the paper is as follows: In section 2, we review
some definitions and properties related to association schemes,
underlying resistor networks of them and corresponding
stratifications. In section 3, the effective resistance in general
resistor networks and underlying resistor networks of association
schemes is reviewed. Section 4 is devoted to calculation of the
effective resistances on underlying resistor networks of association
schemes by using the algebraic combinatoric structures of
corresponding association schemes without using the spectrum of
underlying networks. In section 5, explicit formula for effective
resistances on underlying resistor networks of association schemes
is given in terms of spectrum of underlying networks. The paper is
ended with a brief conclusion and an appendix.
\section{Underlying resistor networks of association schemes }
In this section, we review some preliminary tools about underlying
networks which are considered through this paper. For material not
covered in this section, as well as more detailed information
about association schemes and their underlying graphs, refer to
\cite{Ass.sch.}, \cite{Bose} and \cite{js}.

\textbf{Definition 1} Assume that $V$ and $E$ are vertex and edge
sets of a regular resistor network, respectively (each edge has a
certain conductance). Then, the relations $\{R_i\}_{0\leq i\leq
d}$
on $V\times V$ satisfying the following conditions\\
$(1)\;\ \{R_i\}_{0\leq i\leq d}$ is a partition of $V\times V$\\
$(2)\;\ R_0=\{(\alpha, \alpha) : \alpha\in V \}$\\
$(3)\;\ R_i=R_i^t$ for $0\leq i\leq d$, where
$R_i^t=\{(\beta,\alpha) :(\alpha, \beta)\in R_i\} $\\
$(4)$ For $(\alpha, \beta)\in R_k$, the number  $p^k_{i,j}=\mid
\{\gamma\in X : (\alpha, \gamma)\in R_i \;\ and \;\
(\gamma,\beta)\in R_j\}\mid$ does not depend on $(\alpha, \beta)$
but only on $i,j$ and $k$,\\ define a symmetric association scheme
of class $d$ on $V$ which is denoted by $Y=(V,\{R_i\}_{0\leq i\leq
d})$. Furthermore, if we have $p^k_{ij}=p^k_{ji}$ for all
$i,j,k=0,2,...,d$, then $Y$ is called commutative.

Let $Y=(V,\{R_i\}_{0\leq i\leq d})$ be a commutative symmetric
association scheme of class $d$, then the matrices $A_0,A_1,...,A_d$
defined by
\begin{equation}\label{adj.}
    \bigl(A_{i})_{\alpha, \beta}\;=\left\{\begin{array}{c}
      \hspace{-2.3cm}1 \quad \mathrm{if} \;(\alpha,
    \beta)\in R_i, \\
      0 \quad \mathrm{otherwise} \quad \quad \quad(\alpha, \beta
    \in V) \\
    \end{array}\right.
\end{equation}
are adjacency matrices of $Y$ such that
\begin{equation}\label{ss}
A_iA_j=\sum_{k=0}^{d}p_{ij}^kA_{k}.
\end{equation}
From (\ref{ss}), it is seen that the adjacency matrices $A_0, A_1,
..., A_d$ form a basis for a commutative algebra \textsf{A} known as
the Bose-Mesner algebra of $Y$. This algebra has a second basis
$E_0,..., E_d$ such that
\begin{equation}\label{idem}
E_0 = \frac{1}{N}J, \;\;\;\;\;\;\ E_iE_j=\delta_{ij}E_i,
\;\;\;\;\;\;\ \sum_{i=0}^d E_i=I.
\end{equation}
where, $N:=|V|$ and $J$ is an $N\times N$ all-one matrix in
$\textsf{A}$. The basis $E_i$, for $0\leq i\leq d$ are known as
primitive idempotents of $Y$. Let $P$ and $Q$ be the matrices
relating the two bases for $\textsf{A}$:
$$
A_j=\sum_{i=0}^d P_{ij}E_i, \;\;\;\;\ 0\leq j\leq d,
$$
\begin{equation}\label{m2}
E_j=\frac{1}{N}\sum_{i=0}^d Q_{ij}A_i, \;\;\;\;\ 0\leq j\leq d.
\end{equation}
Then clearly
\begin{equation}\label{pq}
PQ=QP=NI.
\end{equation}
It also follows that
\begin{equation}\label{eign}
A_jE_i=P_{ij}E_i,
\end{equation}
which shows that the $P_{ij}$ (resp. $Q_{ij}$) is the $i$-th
eigenvalue (resp. the $i$-th dual eigenvalue ) of $A_j$ (resp.
$E_j$) and that the columns of $E_i$ are the corresponding
eigenvectors. Thus $m_i=$ rank($E_i$) is the multiplicity of the
eigenvalue $P_{ij}$ of $A_j$ (provided that $P_{ij}\neq P_{kj}$ for
$k \neq i$). We see that $m_0=1, \sum_i m_i=N$, and
$m_i=$trace$E_i=N(E_i)_{jj}$ (indeed, $E_i$ has only eigenvalues $0$
and $1$, so rank($E_k$) equals to the sum of the eigenvalues).

Clearly, each non-diagonal (symmetric) relation $R_i$ of an
association scheme $Y=(V,\{R_i\}_{0\leq i\leq d})$ can be thought of
as the network $(V,R_i)$ on $V$, where we will call it the
underlying network of association scheme $Y$. In other words, the
underlying network $\Gamma=(V,R_1)$ of an association scheme is an
undirected connected network, where the set $V$ and $R_1$ consist of
its vertices and edges, respectively. Obviously replacing $R_1$ with
one of the other relations such as $R_i$, for  $i\neq 0,1$ will also
give us an underlying network $\Gamma=(V,R_i)$ (not necessarily a
connected network) with the same set of vertices but a new set of
edges $R_i$.

An undirected connected network $\Gamma=(V,R_1)$ is called
distance-regular network if the relations are based on distance
function defined as follows: Let the distance between
$\alpha,\beta\in V$ denoted by $\partial(\alpha, \beta)$ is the
length of the shortest walk connecting $\alpha$ and $\beta$ (recall
that a finite sequence $\alpha_0, \alpha_1,..., \alpha_n \in V$ is
called a walk of length $n$ if $\alpha_{k-1}\sim \alpha_k$ for all
$k=1, 2,..., n$, where $\alpha_{k-1}\sim \alpha_k$ means that
$\alpha_{k-1}$ is adjacent with $\alpha_{k}$), then the relations
$R_i$ in distance-regular networks are defined as: $(\alpha,
\beta)\in R_i$ if and only if $\partial(\alpha, \beta)=i$, for
$i=0,1,...,d$, where $d:=$max$\{\partial(\alpha, \beta): \alpha,
\beta\in V \}$ is called the diameter of the network. Since
$\partial(\alpha, \beta)$ gives the distance between vertices
$\alpha$ and $\beta$, $\partial$ is called the distance function.
Clearly, we have $\partial(\alpha, \alpha)=0$ for all $\alpha\in V$
and $\partial(\alpha, \beta)=1$ if and only if $\alpha\sim \beta$.
Therefore, distance-regular networks become metric spaces with the
distance function $\partial$.

In a distance-regular network, we have $p_{j1}^i=0 $ (for $i\neq 0$,
$j$ dose not belong to $\{i-1, i, i+1 \}$), i.e., the non-zero
intersection numbers of the network are given by
\begin{equation}\label{abc}
 a_i=p_{i1}^i, \;\;\;\  b_i=p_{i+1,1}^i, \;\;\;\
 c_i=p_{i-1,1}^i\;\ ,
\end{equation}
respectively (for more details see \cite{jsj}). The intersection
numbers (\ref{abc}) and the valencies $\kappa_i$ satisfy the
following obvious conditions
$$a_i+b_i+c_i=\kappa,\;\;\ \kappa_{i-1}b_{i-1}=\kappa_ic_i ,\;\;\
i=1,...,d,$$
\begin{equation}\label{intersec}
\kappa_0=c_1=1,\;\;\;\ b_0=\kappa_1=\kappa, \;\;\;\ (c_0=b_d=0).
\end{equation}
Thus all parameters of the network can be obtained from the
intersection array $\{b_0,...,b_{d-1};c_1,...,c_d\}$.

By using the equations (\ref{ss}) and (\ref{intersec}), for
adjacency matrices of distance-regular network $\Gamma$, we obtain
$$A_1A_i=b_{i-1}A_{i-1}+(\kappa-b_i-c_i)A_i+c_{i+1}A_{i+1},\;\
i=1,2,...,d-1,$$
\begin{equation}\label{P0}
A_1A_d=b_{d-1}A_{d-1}+(\kappa-c_d)A_d.
\end{equation}
The recursion relations (\ref{P0}), imply that
\begin{equation}\label{P1}
A_i=P_i(A),\;\ i=0,1,...,d.
\end{equation}
\subsection{Stratification} For an underlying network $\Gamma$, let
$W$ denotes the vector space over $C$ consisting of column vectors
whose coordinates are indexed by vertex set $V$ of $\Gamma$, and
whose entries are in $C$ (i.e., $W=C^N$, with $N=|V|$). We observe
that all $N\times N$ matrices with entries from $C$ act on $W$ by
left multiplication. We endow $W$ with the Hermitian inner product
$\langle , \rangle$ which satisfies $\langle u ,v
\rangle=u^t\bar{v}$ for all $u, v \in W$ , where $t$ denotes the
transpose and - denotes the complex conjugation. For all $\beta\in
V$, let $\ket{\beta}$ denote the element of $W$ with a $1$ in the
$\beta$ coordinate and $0$ in all other coordinates. We observe
$\{\ket{\beta} | \beta\in V\}$ is an orthonormal basis for $W$, but
in this basis, $W$ is reducible and can be reduced to irreducible
subspaces $W_i$, $i=0,1,...,d$, i.e.,
\begin{equation}
W=W_0\oplus W_1\oplus...\oplus W_d,
\end{equation}
where, $d$ is diameter of the corresponding association scheme. In
the following we introduce orthonormal basis for irreducible
subspace of $W$ with maximal dimension, explicitly. To do so, for
a given vertex $\alpha\in V$, we define
 $\Gamma_i(\alpha)=\{\beta\in V:
(\alpha, \beta)\in R_i\}$. Then, the vertex set $ V$ can be
written as disjoint union of $\Gamma_i(\alpha)$, i.e.,
 \begin{equation}\label{asso1}
 V=\bigcup_{i=0}^{d}\Gamma_{i}(\alpha).
 \end{equation}
Now, we fix a point $o\in V$ as an origin of the underlying
network, called reference vertex. Then, the relation (\ref{asso1})
stratifies the network into a disjoint union of strata (associate
classes) $\Gamma_{i}(o)$. With each stratum $\Gamma_{i}(o)$ we
associate a unit vector $\ket{\phi_{i}}$ in $W$ (called unit
vector of $i$-th stratum) defined by
\begin{equation}\label{unitv}
\ket{\phi_{i}}=\frac{1}{\sqrt{\kappa_{i}}}\sum_{\alpha\in
\Gamma_{i}(o)}\ket{\alpha},
\end{equation}
where, $\ket{\alpha}$ denotes the eigenket of $\alpha$-th vertex at
the associate class $\Gamma_{i}(o)$ and $\kappa_i=|\Gamma_{i}(o)|$
is called the $i$-th valency of the network
($\kappa_i:=p^0_{ii}=|\{\gamma:(o,\gamma)\in
R_i\}|=|\Gamma_{i}(o)|$). For $0\leq i\leq d$, the unit vectors
$\ket{\phi_{i}}$ of Eq.(\ref{unitv}) form an orthonormal basis for
irreducible submodule of $W$ with maximal dimension denoted by
$W_0$.

Let $A_i$ be the $i$th adjacency matrix of the underlying network
$\Gamma$. From the action of $A_i$ on reference state $\ket{\phi_0}$
($\ket{\phi_0}=\ket{o}$, with $o\in V$ as reference vertex), we have
\begin{equation}\label{Foc1}
A_i\ket{\phi_0}=\sum_{\beta\in \Gamma_{i}(o)}\ket{\beta}.
\end{equation}
 Then by using (\ref{unitv}) and (\ref{Foc1}),
 we obtain
\begin{equation}\label{Foc2}
A_i\ket{\phi_0}=\sqrt{\kappa_i}\ket{\phi_i}.
\end{equation}
By using (\ref{Foc2}), we can write
\begin{equation}\label{Foc3}
A_i\ket{\phi_j}=\frac{1}{\sqrt{\kappa_j}}A_iA_j\ket{\phi_0}=\frac{1}{\sqrt{\kappa_j}}\sum_k
p^k_{ij}A_k\ket{\phi_0}=\frac{1}{\sqrt{\kappa_j}}\sum_k\sqrt{\kappa_k}
p^k_{ij}\ket{\phi_k}.
\end{equation}
Therefore, from the orthonormality of unit vectors $\ket{\phi_i}$,
for $i=0,1,...,d$, we obtain
\begin{equation}\label{Foc4}
\langle
\phi_l|A_i\ket{\phi_j}=\sqrt{\frac{\kappa_l}{\kappa_j}}p^l_{ij}.
\end{equation}
It could be noticed that, in the case of distance-regular networks,
the adjacency matrices $A_i$ are tridiagonal in the basis of
$\ket{\phi_i}$ (see \cite{jss}, for more details).
\section{Effective resistances on resistor networks}
\subsection{General networks}
For a given regular network $\Gamma$ with $N$ vertices and adjacency
matrix $A$, let $r_{ij}=r_{ji}$ be the resistance of the resistor
connecting vertices $i$ and $j$. Hence, the conductance is
$c_{ij}=r^{-1}_{ij}=c_{ji}$ so that $c_{ij}=0$ if there is no
resistor connecting $i$ and $j$. Denote the electric potential at
the $i$-th vertex by $V_i$ and the net current flowing into the
network at the $i$-th vertex by $I_i$ (which is zero if the $i$-th
vertex is not connected to the external world). Since there exist no
sinks or sources of current including the external world, we have
the constraint $\sum_{i=1}^NI_i=0$. The Kirchhoff law states
\begin{equation}\label{resistor}
\sum_{j=1,j\neq i}^Nc_{ij}(V_i-V_j)=I_i,\;\;\  i=1,2,...,N.
\end{equation}
Explicitly, Eq.(\ref{resistor}) reads
\begin{equation}\label{resistor1}
L\vec{V}=\vec{I},
\end{equation}
where, $\vec{V}$ and $\vec{I}$ are $n$-vectors whose components
are $V_i$ and $I_i$, respectively and
\begin{equation}\label{laplas}
L=\sum_{i=1}^Nc_{i}|i\rangle\langle
i|-\sum_{i,j=1}^Nc_{ij}|i\rangle\langle j|
\end{equation}
is the Laplacian of the graph $\Gamma$ with
\begin{equation}
c_i\equiv \sum_{j=1,j\neq i}^Nc_{ij},
\end{equation}
for each vertex $\alpha$. It should be noticed that, $L$ has
eigenvector $(1,1,...,1)^t$ with eigenvalue $0$. Therefore, $L$ is
not invertible and so we define the psudo-inverse of $L$ as
\begin{equation}\label{inv.laplas}
L^{-1}=\sum_{i,\lambda_i\neq0} {\lambda}^{-1}_iE_i,
\end{equation}
where, $E_i$ is the operator of projection onto the eigenspace of
$L^{-1}$ corresponding to eigenvalue $\lambda_i$. It has been shown
that, the effective resistances $R_{\alpha\beta}$ are given by
\begin{equation}\label{eq.res.}
R_{\alpha\beta}=\langle \alpha|L^{-1}|\alpha\rangle+\langle
\beta|L^{-1}|\beta\rangle-\langle \alpha|L^{-1}|\beta\rangle-\langle
\beta|L^{-1}|\alpha\rangle.
\end{equation}
This formula may be formally derived \cite{klein} using Kirchoff 's
laws, and seems to have been long known in the electrical
engineering literature, with it appearing in several texts, such as
Ref. \cite{12}.
\subsection{Underlying resistor networks of association schemes}
In the present paper we deal with special networks which are
underlying networks of some symmetric association schemes. For
these networks, first we choose a vertex, say $\alpha$, as
reference vertex and stratify the network with respect to
$\alpha$. Then, we assume that the conductance between $\alpha$
and $\beta$ is $c_i$ for all $\beta\in \Gamma_i(\alpha)$, i.e.,
the conductances between $\alpha$ and all vertices belonging to
the same strata (with respect to $\alpha$) are the same. Then, the
Laplacian of the underlying network is defined as
\begin{equation}\label{laplas}
L=(\sum_{i=0}^dc_i\kappa_i)I-\sum_{i=0}^dc_iA_i,
\end{equation}
where, $d$ is the diameter of the association scheme. Obviously,
in the case that all nonzero resistances are connecting
resistances and are equal to $1$ ($c\equiv c_1=1$, $c_i=0$ for
$i\neq 1$), the off-diagonal elements of $-L$ are precisely those
of the adjacency matrix $A$ of the network, i.e.,
\begin{equation}\label{laplas1x}
L=\kappa I-A,
\end{equation}
where, $\kappa\equiv\kappa_1=deg(\alpha)$ (in regular networks, the
degree is independent of the vertex $\alpha$). In section $5$, we
will show that in the case of underlying networks of association
schemes, all of the diagonal entries of the pseudo inverse matrix
$L^{-1} $ are equal. By using this result and from the fact that
$L^{-1}$ is a real matrix, the Eq.(\ref{eq.res.}) can be written for
these networks as follows
\begin{equation}\label{eq.res.1}
R_{\alpha\beta}=2(\langle \alpha|L^{-1}|\alpha\rangle-\langle
\alpha|L^{-1}|\beta\rangle).
\end{equation}
\section{Calculating effective resistances on underlying networks
of association schemes without using the spectrum of the networks}
In general, as it will be shown in the next section (the
Eq.(\ref{ress})), in order to calculate the effective resistances on
underlying networks of association schemes, one needs to know the
spectrum of the adjacency matrices $A_i$ for $i=1,...,d$. Although,
in the most cases the spectrum of the Bose-Mesner algebra is known,
but the formula for effective resistances in terms of the spectrum
of the network do not possess a closed form and evaluation of it in
the most cases is not an easy task. In this section, we assume that
all of the conductances except for one of them, say $c\equiv c_1=1$,
are zero and  consider particular underlying networks of association
schemes such that the adjacency matrices $A_i$ are written as
polynomials of the first adjacency matrix $A=A_1$ (not necessarily
of degree $i$). Then, we give a procedure for evaluating the
effective resistances on these networks such that one can calculate
the effective resistances by using the structure of their
Bose-Mesner algebra without any need to know the spectrum of the
adjacency matrices (even if there is no any three-term recursion
relations such as (\ref{P0}) which are satisfied by distance-regular
networks). In the appendix $A$, we show that if the adjacency matrix
of a connected underlying network of association scheme with
diameter $d$ possesses $d+1$ distinct eigenvalues, then all of the
other adjacency matrices $A_i$ for $i=2,...,d$ can be written as
polynomials of $A$.

As it will be shown in section $5$ (Corollary $1$), all of the nodes
$\beta$ belonging to the same stratum with respect to the reference
node $\alpha$, possess the same effective resistance with respect to
$\alpha$. This allows us to write
$$R_{\alpha\beta^{(m)}}=\frac{1}{\kappa_m}\sum_{\beta\in \Gamma_m(\alpha)}R_{\alpha\beta}=\frac{1}{\kappa_m}\sum_{\beta\in V}(A_m)_{\alpha\beta}R_{\alpha\beta}.$$
where, $R_{\alpha\beta^{(m)}}$ denotes the effective resistances
between $\alpha$ and all of the nodes $\beta\in \Gamma_{m}(\alpha)$.
Now, consider underlying networks of association schemes for which
we have $A_m=\sum_{n=0}^{d}c_{mn}A^n$ (recall that, for underlying
networks which satisfy the distance-regularity condition, i.e., the
three-term recursion relations (\ref{P0}) are satisfied, we have
$A_m=P_m(A)=\sum_{n=0}^{m}c_{mn}A^n$, where $P_m$ is a polynomial of
degree $m$). Then, by using (\ref{eq.res.1}) one can obtain
\begin{equation}\label{eq}
R_{\alpha\beta^{(m)}}=\frac{2}{\kappa_m}\sum_{n=0}^{d}c_{mn}[\sum_{\beta\in
V}(A^n)_{\alpha\beta}L^{-1}_{\alpha\alpha}-\sum_{\beta\in
V}(A^n)_{\alpha\beta}L^{-1}_{\alpha\beta}].
\end{equation}
From the fact that, the effective resistances
$R_{\alpha\beta^{(m)}}$ are independent of the choice of the
reference node $\alpha$, one can write
\begin{equation}\label{eq1x}
\sum_{\alpha\in
V}R_{\alpha\beta^{(m)}}=N.R_{\alpha\beta^{(m)}}=\frac{2}{\kappa_m}\sum_{n=0}^{d}c_{mn}[\sum_{\alpha,\beta\in
V}(A^n)_{\alpha\beta}L^{-1}_{\alpha\alpha}-\sum_{\alpha,\beta\in
V}(A^n)_{\alpha\beta}L^{-1}_{\alpha\beta}].
\end{equation}
Now, we note that
$$\sum_{\beta\in
V}(A^n)_{\alpha\beta}=\sum_{\beta,\gamma_1,...,\gamma_{n-1}\in
V}A_{\alpha\gamma_1}A_{\gamma_1\gamma_2}...A_{\gamma_{n-1}\beta}=\kappa^n.$$
Now, assume that all of the conductances $c_i$ equal to zero
except for $c_1\equiv c=1$ which implies that the Eq.
(\ref{laplas1x}) is satisfied. Then, from (\ref{eq1x}), we obtain
$$R_{\alpha\beta^{(m)}}=\frac{2}{N.\kappa_m}\sum_{n=0}^{d}c_{mn}[\kappa^n.tr(L^{-1})-tr(A^nL^{-1})]=\frac{2}{N.\kappa_m}\sum_{n=0}^{d}c_{mn}tr(\frac{\kappa^n\mathbf{1}-A^n}{\kappa\mathbf{1}-A})=$$
\begin{equation}\label{eq1'}
\frac{2}{N.\kappa_m}\sum_{n=1}^{d}c_{mn}tr[(\kappa^{n-1}\mathbf{1}+\kappa^{n-2}A+\kappa^{n-3}A^2+...+A^{n-1})(\mathbf{1}-1/NJ)].
\end{equation}
By using the equality $A^lJ=\kappa^lJ$ (recall that $AJ=\kappa
J$), the Eq.(\ref{eq1'}) can be rewritten as follows
\begin{equation}\label{eq1''}
R_{\alpha\beta^{(m)}}=\frac{2}{N.\kappa_m}\sum_{n=1}^{d}c_{mn}(\sum_{i=1}^n\kappa^{n-i}tr(A^{i-1})-n.\kappa^{n-1}).
\end{equation}
As the above formula indicates, in order to calculate
$R_{\alpha\beta^{(m)}}$, we need to evaluate $tr(A^l)$, for all
$l=1,2,...,d-1$. To this aim, we use the relations
$A_m=\sum_{n=0}^{d}c_{mn}A^n$ for $m=1,...,d$ to write $A^l$ as
$A^l=\sum_{m=0}^{d}c'_{lm}A_m$ and obtain $tr(A^l)=N.c'_{l0}$.
\subsection{Examples}
\textbf{1. Underlying network of association scheme derived from
$Z_5\times Z_5$}\\
In the regular representation, the elements of abelian group
$Z_5\times Z_5$ are written as $S_1^kS_2^l$ with $S_1=S\otimes I$
and  $S_2=I\otimes S$, where $S$ is the shift operator with period
$5$, i.e., $S^5=I_5$. Now, we define the following adjacency
matrices
$$A\equiv A_1=S_1+S_2+S_1S_2+S^4_1+S^4_2+S^4_1S^4_2,$$
$$A_2=S^2_1+S^2_2+S^2_1S^2_2+S^3_1+S^3_2+S^3_1S^3_2,$$
$$A_3=S^3_1+S^3_2+S^3_1S^4_2+S^4_1S^3_2+S^2_1S_2+S_1S^2_2,$$
$$
A_4=S_1S^3_2+S^3_1S_2+S^2_1S^3_2+S^3_1S^2_2+S^2_1S^4_2+S^4_1S^2_2.
$$
Then, one can easily see that the above adjacency matrices
constitute the Bose-Mesner algebra of a symmetric association
scheme. In fact, we have
\begin{equation}\label{hex5}
A^2=6A_0+2A+A_2+2A_3,\;\ AA_2=A+A_2+2A_3+2A_4,\;\
AA_3=2A+2A_2+2A_4,\;\ AA_4=2A_2+2A_3+2A_4.
\end{equation}
The above relations indicate that the underlying network of the
constructed association scheme is not distance regular. By using
(\ref{hex5}), one can evaluate the powers of $A$ as follows
\begin{equation}\label{hex5'}
A^2=6A_0+2A+A_2+2A_3,\;\;\ A^3=12A_0+15A+7A_2+6A_3+6A_4,\;\;\
A^4=90A_0+61A+46A_2+56A_3+38A_4.
\end{equation}
 Then, by solving (\ref{hex5'}) in terms of $A_2,A_3$ and $A_4$, we
 obtain
$$
A_2=\frac{1}{44}(-6A^4+38A^3+54A^2-312A-240A_0),\;\;\;\
A_3=\frac{1}{44}(3A^4-19A^3-5A^2+112A-12A_0),$$
\begin{equation}\label{hex5''}
A_4=\frac{1}{22}(2A^4+9A^3-29A^2+71A+102A_0).
\end{equation}
That is, the coefficients $c_{mn}$ in $A_m=\sum_{n=0}^dc_{mn}A^n$
are given by
$$c_{11}=1 ,\;\ c_{1i}=0\;\ \mbox{for}\;\ i\neq 1 ; \;\;\ c_{20}=\frac{60}{11}, \;\ c_{21}=-\frac{78}{11},\;\ c_{22}=\frac{27}{22},\;\ c_{23}=\frac{19}{22},\;\ c_{24}=-\frac{3}{22}; \;\ c_{30}=-\frac{3}{11},$$
\begin{equation}\label{hex5'''}
c_{31}=\frac{28}{11},\;\ c_{32}=-\frac{5}{44},\;\
c_{33}=-\frac{19}{44},\;\ c_{34}=\frac{3}{44}; \;\
c_{40}=\frac{51}{11},\;\ c_{41}=\frac{71}{22},\;\
c_{42}=-\frac{29}{22},\;\ c_{43}=\frac{9}{22},\;\
c_{44}=\frac{1}{11}.
\end{equation}
Then, by using (\ref{hex5'}), (\ref{hex5'''})  and substituting
$N=25$ and $\kappa=\kappa_2=\kappa_3=\kappa_4=6$ in the result
(\ref{eq1''}), we obtain the effective resistances as follows:
$$R_{\alpha\beta^{(1)}}=\frac{1}{75}c_{11}(25-1)=\frac{24}{75},$$
$$R_{\alpha\beta^{(2)}}=\frac{1}{75}\{24c_{21}+138c_{22}+942c_{23}+5736c_{24})=\frac{112}{275},$$
$$
R_{\alpha\beta^{(3)}}=\frac{1}{75}\{24c_{21}+138c_{22}+942c_{23}+5736c_{24})=\frac{327}{825},$$
\begin{equation}\label{s4new2xx}
R_{\alpha\beta^{(4)}}=\frac{1}{75}\{24c_{21}+138c_{22}+942c_{23}+5736c_{24})=\frac{2942}{275}.
\end{equation}
\textbf{2. Group association scheme $S_4$}\\
In group association schemes, the adjacency matrices are defined
as the class sums of a group in regular representation. For
instance, in the symmetric group $S_4$, the conjugacy classes are
given by
$$
C_0=\{1\},\;\ C_1=\{(12),(13),(14),(23),(24),(34)\},\;\
C_2=\{(123),(132),(124),(142),(134),(143),(234),$$
\begin{equation}\label{cons4}
(243)\},\;\ C_3=\{(12)(34),(13)(24),(1 4)(23)\}, \;\
C_4=\{(1234),(1243),(1324),(1342),(1423),(1432)\}.
\end{equation}
Then, the adjacency matrices are defined as $A_i=\bar{C_i}, \;\
i=0,1,...,4$, i.e., we have
$$A\equiv A_1=(12)+(13)+(14)+(23)+(24)+(34), \;\;\ A_2=(123)+(132)+(124)+(142)+(134)+(143)+(234)+$$
$$(243),\;\ A_3=(12)(34)+(13)(24)+(14)(23),\;\ A_4=(1234)+(1243)+(1324)+(1342)+(1423)+(1432).$$
One can easily show that these adjacency matrices satisfy the
following relations
\begin{equation}\label{s4x}
A^2=6A_0+3A_2+2A_3, \;\;\ AA_2=4A+4A_4, \;\;\ AA_3=A+2A_4, \;\;\
AA_4=4A_2+4A_3,
\end{equation}
above relations indicate that the group association scheme $S_4$  is
not distance regular ( actually group scheme $S_n$ is a
distance-regular one, only for $n=3$). By using (\ref{s4x}), one can
evaluate the powers of $A$ as follows
\begin{equation}\label{s4xx}
A^2=6A_0+3A_2+2A_3,\;\;\ A^3=20A+16A_4,\;\;\ A^4=120
A_0+108A_2+104A_3.
\end{equation}
 Then, by solving (\ref{s4xx}) in terms of $A_2,A_3$ and $A_4$, we
 obtain
\begin{equation}\label{s4xx'}
A_2=-\frac{1}{48}(A^4-52A^2+192A_0),\;\;\
A_3=\frac{1}{32}(A^4-36A^2+96A_0),\;\;\ A_4=\frac{1}{16}(A^3-20A).
\end{equation}
That is, the coefficients $c_{mn}$ in $A_m=\sum_{n=0}^dc_{mn}A^n$
are given by
$$c_{11}=1 ,\;\ c_{1i}=0\;\ \mbox{for}\;\ i\neq 1 ; \;\;\ c_{20}=-4, \;\ c_{21}=0,\;\ c_{22}=\frac{13}{12},\;\ c_{23}=0,\;\ c_{24}=-\frac{1}{48}; \;\ c_{30}=3,$$
\begin{equation}\label{s4new2x}
c_{31}=0,\;\ c_{32}=-\frac{9}{8},\;\ c_{33}=0,\;\
c_{34}=\frac{1}{32}; \;\ c_{40}=0,\;\ c_{41}=-\frac{5}{4},\;\
c_{42}=0,\;\ c_{43}=\frac{1}{16},\;\ c_{44}=0.
\end{equation}
Then, by using (\ref{s4xx}), (\ref{s4new2x})  and substituting
$N=24$ and $\kappa=6,\;\ \kappa_2=8,\;\ \kappa_3=3,\;\ \kappa_4=6$
in the result (\ref{eq1''}), we obtain the effective resistances as
follows:
$$R_{\alpha\beta^{(1)}}=\frac{1}{72}c_{11}(24-1)=\frac{23}{72},\;\;\;\
R_{\alpha\beta^{(2)}}=\frac{1}{96}\{132c_{22}+6048c_{24})=\frac{35}{96},$$
\begin{equation}\label{s4new2xx}
R_{\alpha\beta^{(3)}}=\frac{1}{36}\{132c_{32}+5184c_{34}\}=\frac{3}{8},\;\;\;\
R_{\alpha\beta^{(4)}}=\frac{1}{72}\{23c_{41}+1620c_{43}=\frac{145}{36}.
\end{equation}

In order to give another nontrivial examples of underlying networks
of association schemes which are not distance-regular networks, we
construct two association schemes with diameter $6$ by combining the
class sums of the
symmetric group $S_4$ as follows:\\
\textbf{a)} We define the adjacency matrices as follows:
$$\hspace{-1.5cm}A_0=I,\;\ A\equiv A_1=(12)+(13)+(14),\;\ A_2=(123)+(132)+(124)+(142)+(134)+(143), \;\ A_3=(23)+(24)+(34),$$
\begin{equation}\label{s4a}
\hspace{-1.5cm}A_4=(1234)+(1243)+(1324)+(1342)+(1423)+(1432), \;\;\
A_5=(12)(34)+(13)(24)+(14)(23),\;\ A_6=(234)+(243).
\end{equation}
Then, one can show that the following relations are satisfied
$$A^2=3A_0+A_2,\;\ AA_2=2A+2A_3+A_4,\;\ AA_3=A_2+A_5,$$
\begin{equation}\label{xx}
AA_5=A_3+A_4,\;\ AA_4=A_2+2A_5+3A_6,\;\ AA_6=A_4.
 \end{equation}
 The relations (\ref{xx}) indicate that the underlying network is
not distance-regular (see Eq. (\ref{P0})). Now, in order to evaluate
the effective resistances on this network, we calculate the powers
of the adjacency matrix $A$ as follows:
$$A^2=3A_0+A_2,\;\ A^3=5A+2A_3+A_4,\;\ A^4=15A_0+8A_2+4A_5+3A_6,$$
\begin{equation}\label{s4new}
A^5=31A+20A_3+15A_4,\;\ A^6=93A_0+66A_2+50A_5+45A_6.
\end{equation}
Then by solving the five equations with five unknown
$A_2,...,A_6$, we obtain the following solution
$$A_2=A^2-3A_0,\;\ A_3=\frac{1}{10}(-A^5+15A^3-44A),\;\ A_4=\frac{1}{5}(A^5-10A^3+19A),$$
\begin{equation}\label{s4new1}
A_5=\frac{1}{10}(-A^6+15A^4-54A^2+30A_0),\;\
A_6=\frac{1}{15}(2A^6-25A^4+68A^2-15A_0).
\end{equation}
That is, the coefficients $c_{mn}$ in $A_m=\sum_{n=0}^dc_{mn}A^n$
are given by
$$c_{11}=1 ,\;\ c_{1i}=0\;\ \mbox{for}\;\ i\neq 1 ; \;\;\ c_{20}=-3, \;\ c_{22}=1,\;\ c_{21}=c_{23}=c_{24}=c_{25}=c_{26}=0 ;$$
$$c_{30}=c_{32}=c_{34}=c_{36}=0,\;\ c_{31}=-\frac{22}{5},\;\ c_{33}=\frac{3}{2},\;\ c_{35}=-\frac{1}{10}; \;\ c_{40}=c_{42}=c_{44}=c_{46}=0,$$
$$ c_{41}=\frac{19}{5}, c_{43}=-2,\;\ c_{45}=\frac{1}{5}; \;\ c_{50}=3,\;\ c_{51}=c_{53}=c_{55}=0,\;\ c_{52}=-\frac{27}{5},\;\ c_{54}=\frac{3}{2},\;\ c_{56}=-\frac{1}{10} ;$$
\begin{equation}\label{s4new2}
c_{60}=-1,\;\ c_{61}=c_{63}=c_{65}=0,\;\ c_{62}=\frac{68}{15},\;\
c_{64}=-\frac{5}{3},\;\ c_{66}=\frac{2}{15}.
\end{equation}
Then, by using (\ref{s4new1}), (\ref{s4new2})  and substituting
$N=24$ and $\kappa=3,\;\ \kappa_2=3,\;\ \kappa_3=6,\;\
\kappa_4=2,\;\ \kappa_5=3,\;\ \kappa_6=6$ in the result
(\ref{eq1''}), we obtain the effective resistances as follows:
$$\hspace{-5cm}R_{\alpha\beta^{(1)}}=\frac{1}{36}c_{11}(24-1)=\frac{23}{36},\;\;\ R_{\alpha\beta^{(2)}}=\frac{1}{72}c_{22}(72-6)=\frac{33}{36},$$
$$R_{\alpha\beta^{(3)}}=\frac{1}{36}\{c_{31}(24-1)+c_{33}(\sum_{i=1}^3
3^{3-i}tr(A^{i-1})-27)+c_{35}(\sum_{i=1}^53^{5-i}tr(A^{i-1})-405)\}=\frac{89}{90},$$
$$R_{\alpha\beta^{(4)}}=\frac{1}{36}\{c_{41}(24-1)+c_{43}(\sum_{i=1}^3
3^{3-i}tr(A^{i-1})-27)+c_{45}(\sum_{i=1}^53^{5-i}tr(A^{i-1})-405)\}=\frac{187}{180},$$
$$R_{\alpha\beta^{(5)}}=\frac{1}{36}\{c_{52}(72-6)+c_{54}(\sum_{i=1}^4 3^{4-i}tr(A^{i-1})-108)+c_{56}(\sum_{i=1}^63^{6-i}tr(A^{i-1})-1458)\}=\frac{21}{20},$$
\begin{equation}\label{s4new3}
R_{\alpha\beta^{(6)}}=\frac{1}{24}\{c_{62}(72-6)+c_{64}(\sum_{i=1}^4
3^{4-i}tr(A^{i-1})-108)+c_{66}(\sum_{i=1}^63^{6-i}tr(A^{i-1})-1458)\}=\frac{16}{15}.
\end{equation}
\textbf{b)} If we choose the matrix $A_4$ in (\ref{s4a}) as
adjacency matrix, we obtain another connected network (the matrices
$A_2,A_3,A_5$ and $A_6$ do not define connected networks). Then, one
can obtain
$$A_4A_1=A_2+2A_5+3A_6,\;\ A_4A_2=2A_1+4A_3+3A_4,\;\ A_4A_3=2A_2+2A_5,$$
\begin{equation}
A_4^2=6A_0+3A_2+2A_5+3A_6,\;\ A_4A_5=2A_1+2A_3+A_4,\;\
A_4A_6=2A_1+A_4.
 \end{equation}
which indicate that the underlying network is not distance-regular
(see Eq. (\ref{P0})). Again, in order to evaluate the effective
resistances on this network, we calculate the powers of the
adjacency matrix $A\equiv A_4$ as follows:
$$A^2=6A_0+3A_3+3A+2A_5,\;\ A^3=16(A_1+A_2)+20A_6,\;\ A^4=120A_0+108(A_3+A)+104A_5,$$
\begin{equation}\label{s4new'}
A^5=640(A_1+A_2)+656A_6,\;\ A^6=3936A_0+3888(A_3+A)+3872A_5.
\end{equation}
By solving (\ref{s4new'}), one can write $A_i$ for $i=1,2,3,5,6$ in
terms of powers of $A\equiv A_4$ and (similar to the case $a$)
evaluate effective resistances $R_{\alpha\beta^{(i)}}$ for
$i=1,2,3,5,6$.

It should be noticed that, in distance-regular networks, by using
the three-term recursion relations (\ref{P0}), one can obtain
$$A^2=AA_1=\kappa\mathbf{1}+a_1A+c_2A_2,$$
$$A^3=AA^2=\kappa a_1\mathbf{1}+(\kappa+a_1^2+b_1c_2)A+(a_1+a_2)c_2A_2+c_2c_3A_3,$$
$$A^4=AA^3=I_0\mathbf{1}+I_1A+I_2A_2+I_3A_3+c_2c_3c_4A_4,$$
$$A^5=AA^4=\kappa I_1\mathbf{1}+(I_0+a_1I_1+b_1I_2)A+(c_2I_1+a_2I_2+b_2I_3)A_3+(c_3I_2+a_3I_3+b_3c_2c_3c_4)A_3+$$
\begin{equation}\label{eq1'''}
(c_4I_3+a_4c_2c_3c_4)A_4+c_2c_3c_4c_5A_5,
\end{equation}
where,
$$I_0:=\kappa(\kappa+a_1^2+b_1c_2),\;\ I_1:=a_1(2\kappa+a_1^2+2b_1c_2)+b_1c_2a_2,\;\ I_2:=c_2[\kappa+a_1^2+b_1c_2+a_2(a_1+a_2)+b_2c_3],$$
\begin{equation}
I_3:=c_2c_3(a_1+a_2+a_3).
\end{equation}

Therefore, by using (\ref{eq1''}) and (\ref{eq1'''}), for
distance-regular resistor networks such that all of the conductances
$c_i$ are equal to zero except for one of them, i.e., $c_1\equiv
c=1$ and $c_i=0$ for $i\neq 1$, the effective resistances
$R_{\alpha\beta^{(m)}}$ for $m=1,2,...,5$ are obtained in terms of
the intersection numbers of the network as follows
$$
R_{\alpha\beta^{(1)}}=\frac{2}{\kappa}(\frac{N-1}{N}),
$$
$$
R_{\alpha\beta^{(2)}}=\frac{2}{\kappa
b_1}\{b_1+1-\frac{\kappa+b_1+1}{N}\},
$$$$
R_{\alpha\beta^{(3)}}=\frac{2}{\kappa b_1b_2}\{b_{d-1}c_d+
b_2-\kappa+c_2+b_1b_2-\frac{(\kappa+1)(b_2+c_2)+b_1(\kappa+b_2)}{N}\},
$$
$$
R_{\alpha\beta^{(4)}}=\frac{2}{\kappa
b_1b_2b_3}\{-I_1(1-\frac{1}{N})-\kappa I_2(1-\frac{2}{N})- \kappa
I_3(\kappa+1-3\frac{\kappa}{N})+\kappa^3(1-\frac{4}{N})+\kappa.(\kappa+a_1)\}
,$$
$$
R_{\alpha\beta^{(5)}}=\frac{2}{\kappa
b_1b_2b_3b_4}\{-(I_0+a_1I_1+b_1I_2)(1-\frac{1}{N})-(c_2I_1+a_2I_2+b_2I_3)(\kappa-2)-(c_3I_2+a_3I_3+b_3c_2c_3c_4).$$
\begin{equation}\label{eq1''''}
(\kappa^2(1-\frac{3}{N})+\kappa
)-(c_4I_3+a_4c_2c_3c_4)(\kappa^3(1-\frac{4}{N})+\kappa
(\kappa+a_1))+\kappa^4(1-\frac{5}{N})+\kappa(\kappa^2+\kappa
a_1+\kappa+a_1^2+b_1c_2)\}.
\end{equation}
In Ref. \cite{jsj}, analytical formulas for effective resistances up
to the third stratum, i.e., $R_{\alpha\beta^{(i)}}$ for $i=1,2,3$ on
distance-regular networks have been given by using the stieltjes
function associated with the network, where the results are in
agreement with (\ref{eq1''''}). It should be noticed that, by using
the above result, one can evaluate the limiting value of the
effective resistances in the limit of the large size of the
networks, i.e., in the limit $N\rightarrow \infty$. For instance,
the effective resistances $R_{\alpha\beta^{(1)}}$,
$R_{\alpha\beta^{(2)}}$ and $R_{\alpha\beta^{(3)}}$ tend to
$\frac{2}{\kappa}$, $\frac{2(b_1+1)}{\kappa b_1}$ and
$\frac{2(b_{d-1}c_d+ b_2-\kappa+c_2+b_1b_2)}{\kappa b_1b_2}$,
respectively which are finite for the resistor networks with finite
value of the valency $\kappa$.

In the following, we introduce some interesting distance-regular
networks which are underlying networks of association schemes
derived from symmetric group $S_n$ and its nontrivial subgroups and
calculate the effective resistances on these networks.
\subsubsection{ Association schemes derived from symmetric
group $S_n$} Let $\lambda=(\lambda_1,...,\lambda_m)$ be a partition
of $n$, i.e., $\lambda_1+...+\lambda_m=n$. We consider the subgroup
$S_m\otimes S_{n-m}$ of $S_n$ with $m\leq [\frac{n}{2}]$. Then we
assume the finite set $M^{\lambda}$ (where, the association scheme
is defined on it) as $M^{\lambda}=\frac{S_n}{S_m\otimes S_{n-m}}$
with $|M^{\lambda}|=\frac{n!}{m!(n-m)!}$. In fact, $M^{\lambda}$ is
the set of $(m-1)$-faces of $(n-1)$-simplex (note that, the graph of
an $(n-1)$-simplex is the complete graph with $n$ vertices denoted
by $K_n$). If we denote the vertex $i$ by $m$-tuple
$(i_1,i_2,...,i_m)$, then the relations $R_k$, $k=0,1,...,m$ defined
by
\begin{equation}\label{eq1}
R_k=\{(i,j): \partial(i,j)=k\},\;\;\ k=0,1,...,m,
\end{equation}
where, we mean by $\partial(i,j)$ the number of components that
$i=(i_1,i_2,...,i_m)$ and $j=(j_1,j_2,...,j_m)$ are different
(this is the same as Hamming distance which is defined in coding
theory). It can be shown that, the relations $R_k$, for
$k=0,1,...,m$ define an association scheme on
$\frac{S_n}{S_m\otimes S_{n-m}}$ with diameter $m+1$, where the
adjacency matrices $A_k$, $k=0,1,...,m$ are defined as
\begin{equation}\label{adjsn.}
    \bigl(A_{k})_{i, j}\;=\left\{\begin{array}{c}
      \hspace{-2.65cm}1 \quad \mathrm{if} \;\;\   \partial(i,j)=k, \\
      0 \quad \quad \mathrm{otherwise} \quad \quad \quad  (i, j
    \in M^{\lambda}) \\
    \end{array}\right. , \;\;\ k=0,1,...,m.
\end{equation}
 One should notice that, the representation space $M^{\lambda}$
is a module space which is not irreducible, i.e., $M^{\lambda}$ is
decomposed as
\begin{equation}\label{eq2}
M^{\lambda}\cong \bigoplus_{\mu\unlhd \lambda}S^{\mu},
\end{equation}
where, $\mu\unlhd \lambda$ means that $\mu_1+...+\mu_i\leq
\lambda_1+...+\lambda_i$, for each $i=1,...,m$. The number of
distinct irreducible submodules of $M^{\lambda}$ is $m+1$. The
irreducible submodules $S^{\mu}$ are called Specht modules, where
$S^{\lambda}$ is the same as permutation module. The $m+1$
idempotents are defined by
\begin{equation}\label{eq3}
E_{\mu}=\frac{\chi_{\mu}(e)}{n!}\sum_{g\in
S_n}\chi_{\mu}(g)\rho(g),
\end{equation}
where, $e$ is the identity element, $\chi_{\mu}$ is the character
corresponding to the irreducible submodule $S^{\mu}$ and $\rho$ is
the representation of $S_n$ over $M^{\lambda}$.

For the network with adjacency matrices defined by (\ref{adjsn.}),
one can show that the sizes of strata (valencies) are given by
\begin{equation}\label{stratsn0}
\kappa_0=1,\;\ \kappa_l=\left(\begin{array}{c}
                            m \\
                            m-l
                          \end{array}\right)\left(\begin{array}{c}
                            n-m \\
                            l
                          \end{array}\right) ,\;\;\ l=1,2,...,m
\end{equation}
(clearly we have $N=\sum_{l=0}^{m}\kappa_l=\left(\begin{array}{c}
                            n \\
                            m
                          \end{array}\right)=\frac{n!}{m!(n-m)!}=|M^{\lambda}|$). If we stratify the
network with respect to reference node
$\ket{\phi_0}=\ket{i_1,i_2,...,i_m}$, the unit vectors
$\ket{\phi_i}$, $i=1,...,m$ are defined as
$$\ket{\phi_1}=\frac{1}{\sqrt{\kappa_1}}(\sum_{i'_1\neq i_1}\ket{i'_1,i_2,...,i_m}+\sum_{i'_2\neq i_2}\ket{i_1,i'_2,i_3,...,i_m}+...+\sum_{i'_m\neq i_m}\ket{i_1,...,i_{m-1},i'_m}),$$
$$\ket{\phi_2}=\frac{1}{\sqrt{\kappa_2}}\sum_{k\neq l=1}^m\sum_{i'_l\neq i_l;i'_k\neq i_k}\ket{i_1,...i_{l-1},i'_l,i_{l+1},...,i_{k-1},i'_k,i_{k+1}...,i_m},$$
$$\vdots$$
\begin{equation}\label{unitvsn}
\ket{\phi_m}=\frac{1}{\sqrt{\kappa_m}}\sum_{i'_{1}\neq
i_{1};...;i'_{m}\neq i_{m}}\ket{i'_1,i'_2,...,i'_{m}}.
\end{equation}
The constructed network as in the above is a distance-regular
network with intersection array as follows
\begin{equation}\label{intsn}
b_l=(m-l)(n-m-l)\;\  ;  \;\;\ c_l=l^2.
\end{equation}
Then, by using the Eq. (\ref{P0}), one can obtain
\begin{equation}\label{QDRsn}
AA_l=(m-l+1)(n-m-l+1)A_{l-1}+l(n-2l)A_l+(l+1)^2A_{l+1}.
\end{equation}
In the following, we consider the case $m=2$, where the vertices are
edges of a complete graph $K_n$ in details and calculate the
effective resistances.

In the case of $m=2$, we have three kinds of relations as follows
$$R_0=\{((ij),(ij))\},\;\;\;\ R_1=\{((ij),(ik)),((ij),(kj)):j\neq k,i\neq k\},$$
\begin{equation}\label{eq4}
R_2=\{((ij),(kl)):i\neq k,j\neq l\},
\end{equation}
for $i<j=1,2,...,n; \;\  k<l=1,...,n$. Therefore, we have three
adjacency matrices $A_0,A_1\equiv A$ and $A_2$, where
$A_0=I_{n(n-1)/2}$ and
$$(A)_{ij,kl}=\delta_{ik}(1-\delta_{jl})+\delta_{jl}(1-\delta_{ik}),$$
\begin{equation}\label{adj}
(A_2)_{ij,kl}=(1-\delta_{ik})(1-\delta_{jl}),\;\;\
i(k)<j(l)=1,2,...,n.
\end{equation}
For a given vertex $\ket{ij}$, $i<j$ as reference vertex, the
stratification basis $\{\ket{\phi_i}\}_{i=0,1,2}$ defined by
(\ref{unitv}), are obtained as
$$\ket{\phi_0}=\ket{ij},\;\;\ i<j=1,...,n,$$
$$\ket{\phi_1}=\frac{1}{\sqrt{2(n-2)}}(\sum_{k\neq j=1}^n\ket{ik}+\sum_{k\neq
i=1}^n\ket{kj}),$$
\begin{equation}\label{strata}
\ket{\phi_2}=\frac{1}{\sqrt{\frac{(n-2)(n-3)}{2}}}\sum_{l\neq
m=1\neq j\neq k}^n\ket{lm}.
 \end{equation}
Then, by using (\ref{adj}), one can obtain
$$
A\ket{\phi_0}=\sqrt{2(n-2)}\ket{\phi_1},
$$
$$
A\ket{\phi_1}=\sqrt{2(n-2)}\ket{\phi_0}+(n-2)\ket{\phi_1}+2\sqrt{(n-3)}\ket{\phi_2},
$$
\begin{equation}\label{tri2}
A\ket{\phi_2}= 2\sqrt{n-3}\ket{\phi_1}+2(n-4)\ket{\phi_2}.
\end{equation}
By using (\ref{stratsn0}) and (\ref{intsn}), we have
\begin{equation}\label{int}
\kappa_0=1,\;\ \kappa=\kappa_1=2(n-2),\;\
\kappa_2=\frac{(n-2)(n-3)}{2}; \;\;\
\{b_0,b_1;c_1,c_2\}=\{2(n-2),n-3;1,4\}.
\end{equation}
Then, by using the recursion relations (\ref{QDRsn}), one can
write
$$A^2=2(n-2). I_{\frac{n(n-1)}{2}}+(n-2).A+4A_2,$$
\begin{equation}\label{tri3}
AA_2=(n-3)A+2(n-4)A_2.
\end{equation}
Now, by using the result (\ref{eq1''''}), the effective resistances
are evaluated as follows
\begin{equation}
R_{\alpha\beta^{(1)}}=\frac{n(n-1)-2}{n(n-1)(n-2)},\;\;\
R_{\alpha\beta^{(2)}}=\frac{n(n-1)+6}{n(n-1)(n-3)}.
\end{equation}
\section{Calculating effective resistances
on underlying networks of association schemes by using the spectrum
of the networks} In the following, we use the algebraic combinatoric
structures of underlying resistor networks of association schemes in
order to calculate effective resistances in (\ref{eq.res.}) in terms
of corresponding association scheme's parameters. By using
(\ref{m2}), the Laplacian (\ref{laplas}) can be written as
\begin{equation}\label{laplas00}
L=\sum_{k=0}^d(\sum_{i=0}^dc_i(\kappa_i-P_{ik}))E_k.
\end{equation}
Then, we have
\begin{equation}\label{laplas1}
L^{-1}=\sum_{k=1}^d\frac{E_k}{\sum_{i=0}^dc_i(\kappa_i-P_{ik})}
\end{equation}
Now, for each $\alpha$ and $\beta$, we consider $\alpha$ as
reference vertex and $\beta\in \Gamma_l(\alpha)$. Then, the
diagonal entries of $L^{-1}$ are all the same and equal to
\begin{equation}\label{Laa}
L^{-1}_{\alpha\alpha}=\sum_{k=1}^d\frac{\langle\alpha|E_k|\alpha\rangle}{\sum_{i=0}^dc_i(\kappa_i-P_{ik})}
=\frac{1}{N}\sum_{k=1}^d\frac{m_k}{\sum_{i=0}^dc_i(\kappa_i-P_{ik})},
\end{equation}
where, we have used the fact that
$\langle\alpha|E_k|\alpha\rangle=\frac{m_k}{N}$ with
$m_k=rank(E_k)$. Also, we have
$$L^{-1}_{\beta\alpha}=\frac{1}{\sqrt{\kappa_l}}\langle\phi_l|L^{-1}|\alpha\rangle=\frac{1}{\kappa_l}\langle\alpha|A_lL^{-1}|\alpha\rangle
=\frac{1}{\kappa_l}\langle\alpha|\sum_{k=1}^d\frac{A_lE_k}{\sum_{i=1}^dc_i(\kappa_i-P_{ik})}|\alpha\rangle$$
\begin{equation}\label{ress1}
=\frac{1}{\kappa_l}\sum_{k=1}^d\frac{P_{lk}\langle\alpha|E_k|\alpha\rangle}{\sum_{i=1}^dc_i(\kappa_i-P_{ik})}
=\frac{1}{N\kappa_l}\sum_{k=1}^d\frac{P_{lk}m_k}{\sum_{i=1}^dc_i(\kappa_i-P_{ik})}.
\end{equation}
Therefore, by using (\ref{eq.res.}), we obtain our main result as
\begin{equation}\label{ress}
R_{\alpha\beta^{(l)}}=\frac{2}{N\kappa_l}\sum_{k=1}^d
\frac{m_k(\kappa_l-P_{lk})}{\sum_{i=1}^dc_i(\kappa_i-P_{ik})}\;\
,\;\;\;\ \forall \;\ \beta\in \Gamma_l(\alpha).
\end{equation}
As the result (\ref{ress}) indicates, in order to calculate the
effective resistances on underlying networks of association schemes,
one needs to know the spectrum of the adjacency matrices $A_i$ for
$i=1,...,d$, i.e., one needs to know $P_{il}$ for $i,l=0,...,d$. It
should be also noticed that, in general the sums appearing in the
Eq. (\ref{ress}) do not possess closed form and evaluation of them
in the most cases is not an easy task. Despite of these problems,
the result (\ref{ress}) leads us to some
important facts. For instance,  we conclude the following corollaries from the result (\ref{ress}):\\
\textbf{Corollary 1} The effective resistances between a given node
$\alpha$ and all of the nodes $\beta$ belonging
to the same strata with respect to $\alpha$ are the same.\\
\textbf{Corollary 2} For the networks such that all of the
conductances $c_i$ are equal to zero except for one of them, i.e.,
$c_1\equiv c=1$ and $c_i=0$ for $i\neq 1$, we have
\begin{equation}\label{foster}
\sum_{_{\alpha,\beta; \beta\sim
\alpha}}R_{\alpha\beta}=\frac{N\kappa_1}{2}R_{\alpha\beta^{(1)}}=\sum_{k=1}^d
\frac{m_k(\kappa_l-P_{lk})}{(\kappa_l-P_{lk})}=\sum_{k=1}^d m_k=N-1.
\end{equation}
In fact, this particular result is true for any $N$-site connected
network and was long ago established by Foster \cite{foster} and by
Weinberg \cite{Wein}. In Ref. \cite{klein1}, the sum in
(\ref{foster}) has been identified as one of the sum rules of
effective resistance as a metric (called resistance-distance) which
is an invariant of the graph. It should be noticed that, the result
(\ref{foster}) is special case of the following general result\\
\textbf{Theorem} For any $N$-site underlying network of association
scheme, we have
\begin{equation}\label{foster1}
S=1/2\sum_{_{\alpha,\beta}}\mathcal{A}_{\alpha\beta}R_{\alpha\beta}=N-1,
\end{equation}
where, $\mathcal{A}:=\sum_{i}c_iA_i$.\\
\textbf{proof.} By using the Eq.(\ref{eq.res.1}) and denoting
$C=\sum_{l}c_l\kappa_l$, we have
$$
S=1/2\sum_{_{\alpha,\beta}}\mathcal{A}_{\alpha\beta}R_{\alpha\beta}=\sum_{_{\alpha,\beta}}\mathcal{A}_{\alpha\beta}(C\mathbf{1}-\mathcal{A})^{-1}_{\alpha\alpha}-\sum_{_{\alpha,\beta}}\mathcal{A}_{\alpha\beta}(C\mathbf{1}-\mathcal{A})^{-1}_{\alpha\beta}=
$$
\begin{equation}\label{foster2}
\underbrace{\sum_{l}c_l\kappa_l}_{C}.
tr(\frac{1}{C\mathbf{1}-\mathcal{A}})-tr(\frac{\mathcal{A}}{C\mathbf{1}-\mathcal{A}})=tr(\frac{C\mathbf{1}-\mathcal{A}}{C\mathbf{1}-\mathcal{A}})=tr(\mathbf{1}-1/NJ)=N-1.
\end{equation}
\textbf{Corollary 3} By using the result (\ref{foster1}), one can
obtain a linear dependance between effective resistances as follows
\begin{equation}\label{foster1'}
c_1\kappa_1R_{\alpha\beta^{(1)}}+c_2\kappa_2R_{\alpha\beta^{(2)}}+...+c_d\kappa_dR_{\alpha\beta^{(d)}}=\frac{2(N-1)}{N}.
\end{equation}
\textbf{proof.} For a given $\alpha$, the conductance between
$\alpha$ and all of the nodes $\beta$ which have the relation $l$
with $\alpha$ ($\beta\in \Gamma_l(\alpha)$), is $c_l$ (the number
of such nodes $\beta$ is equal to $\kappa_l$). Then, by using
(\ref{foster1}) one can write
\begin{equation}\label{foster2}
\frac{1}{2}\sum_{_{\alpha,\beta}}\mathcal{A}_{\alpha\beta}R_{\alpha\beta}=\frac{1}{2}N(c_1\kappa_1R_{\alpha\beta^{(1)}}+c_2\kappa_2R_{\alpha\beta^{(2)}}+...+c_d\kappa_dR_{\alpha\beta^{(d)}})=N-1.
\end{equation}
Clearly, if only one of the conductances ($c_1\equiv c$) be non
zero, we will have
\begin{equation}\label{foster3}
R_{\alpha\beta^{(1)}}=\frac{2(N-1)}{Nc\kappa_1},
\end{equation}
which is the same as the result (\ref{foster}).

One should notice that, for most underlying networks of association
schemes the combinatorics structures of the networks such as $m_i$
(the rank of the idempotents), $\kappa_i$ (the valency of adjacency
matrix $A_i$) and first eigenvalue matrix $P$, are known. For
example, for all underlying networks of symmetric group association
schemes, where the adjacency matrices are class sums of the group,
these properties are easily evaluated. In fact, for these networks
we have
\begin{equation}\label{group}
m_i=d^2_i,\;\;\
\kappa_i=\frac{|C_i|^2}{|G|}\sum_{\chi}\chi^2(g_i),\;\
\mbox{with}\;\ g_i\in C_i,\;\;\
P_{il}=\frac{\kappa_l}{d_i}\chi_i(\alpha_l),
\end{equation}
where, $\chi_i$ is the character of the $i$-th irreducible
representation of the group $G$, $d_i=\chi_i(0)$ and $C_i$ is the
$i$-th conjugacy class of $G$. Then, by using the Eq. (\ref{ress}),
one can obtain
\begin{equation}\label{ressgroup}
R_{\alpha\beta^{(l)}}=\frac{2}{|G|}\sum_{k=1}^d
\frac{d_k(d_k-\chi_k(g_l))}{\sum_{i=1}^dc_i\kappa_i(1-\frac{\chi_k(g_i)}{d_k})}\;\
,\;\;\;\ \forall \;\ \beta\in \Gamma_l(\alpha),
\end{equation}

Although the result (\ref{ress}) can be applied to all underlying
networks of association schemes such as distance-regular and
strongly regular networks \cite{Ass.sch.} and underlying networks of
QD and GQD types \cite{js,jss} where the corresponding combinatorics
structures can be evaluated easily, in the following we will
consider only special underlying networks of association schemes
which we construct by using the orbits of the point groups
corresponding to finite lattices such that in the limit of the large
size of the lattices, we obtain root lattices of type $A_n$ (for
more details see Ref. \cite{rootlatt}).
\subsection{Examples} In
this section, we consider some examples of the underlying networks
of association schemes such as cycle network, infinite line network,
hypercube network, finite and infinite square lattice and hexagonal
network.
\subsubsection{Cycle network $C_{2k}$} The cycle network
$C_{N}$ with $N=2k$ vertices is a simple example which is underlying
network of association scheme of class $k$, where all conductances
$c_i$ for $i\neq 1$ are zero and $c_1\equiv c=1$. The adjacency
matrices are given by
\begin{equation}\label{adj1}
A_0=1,\;\ A_i=S^i+S^{-i}\;\ \mbox{for}\;\ i=1,...,k-1,\;\
\mbox{and}\;\ A_k=S^k,
\end{equation}
where, $S$ is the shift operator with period $N$, i.e., $S^{N}=I$.
The idempotents are easily written as
\begin{equation}\label{idem}
E_0=\frac{1}{N}J,\;\ E_i=\ket{\tilde{i}}\langle
\tilde{i}|+\ket{\tilde{-i}}\langle \tilde{-i}|,\;\ \mbox{for}\;\
i=1,...,k-1,\;\ E_k=\ket{\tilde{k}}\langle \tilde{k}|,
\end{equation}
where
\begin{equation}
\ket{\tilde{i}}:=\frac{1}{\sqrt{N}}\left(\begin{array}{c}
                                          1 \\
                                          \omega^i \\
                                          \vdots \\
                                          \omega^{i(N-1)} \\
                                        \end{array}\right).
\end{equation}
Therefore, we have $m_0=m_k=1$, $m_i=2$, for $i=1,...,k-1$. Clearly,
we have $|\Gamma_i(\alpha)|=2$, for $i=1,...,k-1$ and
$|\Gamma_k(\alpha)|=1$. Also, from (\ref{adj1}), it can be seen that
the spectrum of $A_i$, is given by
\begin{equation}\label{spect}
P_{il}=2\cos\frac{2\pi il}{N},\;\;\ l=0,1,...,k
\end{equation}

Now, by using (\ref{ress}), we obtain
\begin{equation}\label{cycres}
R_{\alpha\beta^{(l)}}=\frac{1}{N}\sum_{i=1}^k\frac{m_i(1-\cos
\frac{2\pi il}{N})}{1-\cos\frac{2\pi i}{N}}=
\frac{1}{N}(2\sum_{i=1}^{k-1}\frac{1-\cos\frac{2\pi
il}{N}}{1-\cos\frac{2\pi i}{N}}+\frac{1-(-1)^l}{2}).
\end{equation}
 For example, if $\beta\in \Gamma_1(\alpha)$ ($l=1$), we have
\begin{equation}
R_{\alpha\beta^{(1)}}=\frac{1}{N}(2(\frac{N}{2}-1)+1)=\frac{(N-1)}{N}.
\end{equation}

In the limit of the large $N$, the cyclic network $C_{2k}$ tend to
the infinite line network which is the same as the root lattice
$A_1$. In this case, the eigenvalues $P_{il}$ tend to $\lambda=2\cos
lx$ and the equation (\ref{cycres}) is replaced with
\begin{equation}\label{int}
R_{\alpha\beta^{(l)}}=\frac{1}{2\pi}\int_{0}^{2\pi}\frac{1-\cos
lx}{1-\cos x}dx.
\end{equation}
The integral (\ref{int}) can be evaluated by the method of residues
and gives the following simple result
\begin{equation}\label{int1}
R_{\alpha\beta^{(l)}}=l,\;\;\ \mbox{for all}\;\ \beta\in
\Gamma_l(\alpha).
\end{equation}
\subsubsection{Hypercube network}
For Hypercube network (known also as binary Hamming scheme denoted
by $H(n,2)$) with $N=2^n$ vertices, the adjacency matrices are given
by
\begin{equation}
A_i=\sum_{perm.}\underbrace{\sigma_x\otimes\sigma_x...\otimes\sigma_x}_{i}
\underbrace{\otimes I_2\otimes...\otimes I_2}_{n-i},\;\ i=0,1,...,n,
\end{equation}
where, the summation is taken over all possible nontrivial
permutations. In fact, the underlying network is the cartesian
product of $n$-tuples of complete network $K_2$. Also it can be
shown that, the idempotents $\{E_0,E_1,...,E_n\}$ are symmetric
product of $n$-tuples of corresponding idempotents of complete
network $K_2$. That is, we have
\begin{equation}
E_i=\sum_{perm.}\underbrace{E_-\otimes E_-...\otimes E_-}_{i}
\underbrace{\otimes E_+\otimes...\otimes E_+}_{n-i},\;\;\
i=0,1,...,n,
\end{equation}
where
\begin{equation}
E_{\pm}=\frac{1}{2}(I\pm\sigma_x).
\end{equation}
It is well known that, the eigenvalues $P_{il}$ are given by
\begin{equation}
P_{il}=K_l(i),
\end{equation}
where $K_l(x)$ are the Krawtchouk polynomials defined as
\begin{equation}
K_l(x)=\sum_{i=1}^{l}
 \left(\begin{array}{c}
   x \\
   i \\
       \end{array}\right)\left(\begin{array}{c}
   n-x \\
   l-i \\
       \end{array}\right)(-1)^i.
\end{equation}
Also, we have
\begin{equation}
\kappa_i=m_i=\frac{n!}{i!(n-i)!}.
\end{equation}
Therefore, by using (\ref{ress}) we obtain effective resistance
between two arbitrary nodes $\alpha$ and $\beta$, such that
$\beta\in \Gamma_l(\alpha)$ as
\begin{equation}
R_{\alpha\beta^{(l)}}=\frac{2l!}{2^nn(n-1)...(n-l+1)}\sum_{k=1}^n\frac{n!(\frac{n(n-1)...(n-l+1)}{l!}-K_k(l))}{k!(n-k)!\sum_{i=1}^nc_i(\frac{n!}{i!(n-i)!}-K_k(i))}
\end{equation}
For $n=2$ (square) and $l=1$, we have
\begin{equation}
R_{\alpha\beta^{(1)}}=\frac{1}{4}\sum_{i=1}^2\frac{2(2-K_i(1))}{i!(2-i)!(2(c_1+c_2)-c_1K_i(1)-c_2K_i(2))}=\frac{1}{4}\frac{3c_1+c_2}{c_1(c_1+c_2)},
\end{equation}
where, for $c_2=0$ and $c\equiv c_1=1$, we obtain the simple result
\begin{equation}
R_{\alpha\beta^{(1)}}=\frac{3}{4c}=\frac{3}{4}.
\end{equation}
\subsubsection{$d$-dimensional periodic networks} In this subsection
we consider two examples of the networks which are underlying
networks of association schemes constructed from finite root
lattices of type $A_2$ and $A_1\times A_1$ (for more details see
\cite{rootlatt}). These networks are known as hexagonal and square
lattices, respectively. To this aim, first we briefly recall some of
the main facts about the root lattices of type $A_n$.\\
\textbf{a) Root lattices of type $A_n$}\\ It is well known that a
Coxeter-Dynkin diagram determines a system of simple roots in the
Euclidean space $E_n$. The finite group $W$, generated by the
reflections through the hyperplanes perpendicular to roots
$\alpha_i$, $i=1,...,n$
\begin{equation}\label{wyle}
r_{i}(\beta)=\beta-2\frac{(\alpha_i,\beta)}{(\alpha_i,\alpha_i)}\alpha_i\in
R,
\end{equation}
is called a Weyl group (for the theory of such groups, see
\cite{wyle} and \cite{wyle1}). An action of elements of the Weyl
group $W$ upon simple roots leads to a finite system of vectors,
which is invariant with respect to $W$. A set of all these vectors
is called a system of roots associated with a given Coxeter-Dynkin
diagram (for a description of the correspondence between simple Lie
algebras and Coxeter-Dynkin diagrams, see, for example,
\cite{coxter}). It is proven that roots of $R$ are linear
combinations of simple roots with integral coefficients. Moreover,
there exist no roots which are linear combinations of simple roots
$\alpha_i$, $i=1,2,...,n$, both with positive and negative
coefficients. The set of all linear combinations
\begin{equation}
Q=\{\sum_{i=1}^n a_i\alpha_i\;\ |\;\ a_i \in Z\}\equiv \bigoplus_i
Z\alpha_i,
\end{equation}
 is called a root lattice corresponding
to a given Coxeter-Dynkin diagram. Root system $R$ which corresponds
to Coxeter-Dynkin diagram of Lie algebra of the group $SU(n+1)$,
gives root lattice $A_n$. For example root system $A_2$
(corresponding to lie algebra of SU(3)), where the roots form a
regular hexagon and $\alpha$ and $\beta$ are simple roots. This
lattice is sometimes called hexagonal lattice or triangular lattice.

It is convenient to describe root lattice $A_n$ and its Weyl group
in the subspace of the Euclidean space $E_{n+1}$, given by the
relation $x_1+x_2+...+x_{n+1}=0$, where $x_1$, $x_2$,..., $x_{n+1}$
are the orthogonal coordinates of a point $x\in E_{n+1}$. The unit
vectors in directions of these coordinates are denoted by $e_j$,
respectively. Clearly, $e_i\perp e_j$, $i\neq j$. The set of roots
is given by the vectors $ \alpha_{ij}=e_i-e_j$ for $i\neq j$. The
roots $\alpha_{ij}$, with $i<j$ are positive and the roots
$\alpha_i\equiv \alpha_{i,i+1}=e_i-e_{i+1}$ for $i=1,...,n$,
constitute the system of simple roots.

Now, recall that the point group corresponding to a lattice is a
group of geometric symmetries leaving a point of the lattice fixed.
For any root lattice, the point group is the same as the group of
all automorphisms of the root system (i.e., the group of all
isomorphisms of the root system onto itself). Then, the Weyl group
is a normal subgroup of this group of automorphisms \cite{coxter}.
It has been shown that \cite{coxter}, the point group of the root
lattices is equal to the semidirect group $W\rtimes S_{_{C-D}}$,
where $W$ is the corresponding Weyl group and $S_{_{C-D}}$ is the
group of the symmetries of the Coxeter-Dynkin diagram of the lattice
(all automorphisms which map the Coxeter-Dynkin diagram to itself).
Then, one can see that the point group of the root lattice $A_n$ is
the semidirect product group $S_{n+1}\rtimes Z_2$, where $S_{n+1}$
is the symmetric group (for more details see
\cite{rootlatt,coxter}).

Now, consider a $d$-dimensional lattice, periodic in each direction
with period $m$ and total number of $N=m^d$ vertices. Each vertex of
the lattice corresponds to a basis state $\ket{a}$, where $a$ is a
$d$-component vector with components $a_1,...,a_d\in
\{0,1,...,m-1\}$. In the limit of the large $m$, this lattice tend
to $\underbrace{A_1\times...\times A_1}_d$, the cartesian product of
$d$ root lattices $A_1$. In the following we show that, this lattice
is an underlying network of association scheme which can be derived
from finite abelian group $Z^{\otimes
d}_m=\underbrace{Z_m\times...\times Z_m}_d$ ($m\geq3$). Therefore,
we can obtain effective resistances in terms of scheme's properties.

The generators of the lattice $\underbrace{A_1\times...\times
A_1}_d$ are $S_1$,..., $S_d$ where $S_i=I\otimes...\otimes I\otimes
\underbrace{S}_i\otimes I\otimes...\otimes I$ with $S^m=I$. Then,
one can show that the point group of the lattice is $Z^{\otimes
d}_2\rtimes \mathbf{S}_d$, where $Z^{\otimes d}_2$ is the
corresponding Wyle group generated by the reflections
$S_1\rightarrow S^{-1}_1; S_2\rightarrow S^{-1}_2; ...;
S_d\rightarrow S^{-1}_d$, and $\mathbf{S}_d$ is the symmetry group
which contains all possible permutations of the simple roots $S_1,
S_2,...,S_d$ (recall that, this is the same as the symmetries of the
corresponding Coxeter-Dynkin diagram of the lattice which consists
of the $d$ disconnected simple roots).

Then, the adjacency matrix of the underlying network (which is the
same as the orbit $O(S_1)$) is
\begin{equation}
A=S_1+...+S_d+S^{-1}_1+...+S^{-1}_d.
\end{equation}
In fact, the orbits of the point group corresponding to the lattice
form a partition $P=\{P_{i},i=(i_1,...,i_d)\}$ for $Z^{\otimes
d}_m$. Then, the adjacency matrices $A_{i}$ are defined as the sum
of all elements of $P_{i}$ in the regular representation, i.e., we
define
\begin{equation}
A_{i}=\sum_{g\in P_{i}} g,
\end{equation}
More clearly, one can see that
$$A_{_{\mathbf{i}=(i_1,i_2,...,i_d)}}=O(S^{i_1}_1S^{i_2}_2...S^{i_d}_d)=S^{i_1}_1S^{i_2}_2...S^{i_d}_d+\small{\mathrm{perm.}}+S^{-i_1}_1S^{i_2}_2...S^{i_d}_d+\small{\mathrm{perm}}+S^{i_1}_1S^{-i_2}_2S^{i_3}_3...S^{i_d}_d+\small{\mathrm{perm}.}+$$
\begin{equation}\label{adjd}
...+S^{i_1}_1S^{i_2}_2...S^{i_{d-1}}_{d-1}S^{-i_d}_d+\small{\mathrm{perm.}}+S^{-i_1}_1S^{-i_2}_2S^{i_3}_3...S^{i_d}_d+\small{\mathrm{perm}.}+...+S^{-i_1}_1S^{-i_2}_2...S^{-i_d}_d+\small{\mathrm{perm.}}
.
\end{equation}
where, the ``perm." after each term, denotes all permutations of the
indices $1,2,...,d$ in that term. From Eq.(\ref{adjd}), it can be
easily seen that for these networks, the spectrum of the adjacency
matrices can be find easily, because the adjacency matrices are
diagonalized by Fourier matrix $F_m\otimes...\otimes F_m$,
simultaneously. The corresponding idempotents are given by
$$E_{_{\mathbf{i}=(i_1,i_2,...,i_d)}}=E_{i_1}\otimes E_{i_2}\otimes...\otimes E_{i_d}+\small{\mathrm{perm.}}+E_{-i_1}\otimes E_{i_2}\otimes...\otimes E_{i_d}+\small{\mathrm{perm}}+E_{i_1}\otimes E_{-i_2}\otimes E_{i_3}\otimes...\otimes E_{i_d}+\small{\mathrm{perm}.}+$$
\begin{equation}\label{idemd}
...+E_{i_1}\otimes E_{i_2}\otimes...\otimes E_{i_{d-1}}\otimes
E_{-i_d}+\small{\mathrm{perm.}}+E_{-i_1}\otimes E_{-i_2}\otimes
E_{i_3}\otimes...\otimes
E_{i_d}+\small{\mathrm{perm}.}+...+E_{-i_1}\otimes
E_{-i_2}\otimes...\otimes E_{-i_d}+\small{\mathrm{perm.}},
\end{equation}
where $E_i=|i\rangle\langle i|$ with
$|i\rangle=\frac{1}{\sqrt{m}}(1,\omega,..., \omega^{m-1})^t$ for
$i=0,1,...,m-1$ (see Eq.(\ref{idem})). From (\ref{adjd}), one can
deduce that the eigenvalues of the adjacency matrices
$A_{\mathbf{i}}$ are given by
$$\lambda^{(l_1,...,l_d)}_{_{\mathbf{i}=(i_1,i_2,...,i_d)}}=2[\cos\frac{2\pi(l_1i_1+...+l_di_d)}{m}+\small{\mathrm{perm}.}+\cos\frac{2\pi(-l_1i_1+l_2i_2...+l_di_d)}{m}+\small{\mathrm{perm}.}+$$
$$
...+\cos\frac{2\pi(l_1i_1+...+l_{d-1}i_{d-1}-l_di_d)}{m}+\small{\mathrm{perm}.}+\cos\frac{2\pi(-l_1i_1-l_2i_2+l_3i_3+...+l_di_d)}{m}+\small{\mathrm{perm}.}+...+$$
\begin{equation}\label{eigd}
\cos\frac{2\pi(-l_1i_1-l_2i_2-...-l_{n}i_{n}+l_{n+1}i_{n+1}+...+l_di_d)}{m}+\small{\mathrm{perm}.}],\;\;\
n:=\lfloor d/2\rfloor.
\end{equation}

In the limit of the large size of the lattice, the eigenvalues tend
to
$$\lambda^{(l_1,...,l_d)}(x_1,...,x_d)=2[\cos(l_1x_1+...+l_dx_d)+\small{\mathrm{perm}.}+\cos(-l_1x_1+l_2x_2...+l_dx_d)+\small{\mathrm{perm}.}+$$
$$
...+\cos(l_1x_1+...+l_{d-1}x_{d-1}-l_di_d)+\small{\mathrm{perm}.}+\cos(-l_1x_1-l_2x_2+l_3x_3+...+l_dx_d)+\small{\mathrm{perm}.}+...+$$
\begin{equation}\label{eigdinf}
\cos(-l_1x_1-l_2x_2-...-l_{n}x_{n}+l_{n+1}x_{n+1}+...+l_dx_d)+\small{\mathrm{perm}.}],
\end{equation}
where, $x_k=\lim_{{i_k,m}\rightarrow \infty}2\pi i_k/m$ for
$k=1,2,...,d$. By assumption of $c_1\equiv c=1$ and $c_i=0$ for all
$i\neq 1$ and using the fact that $\upsilon_1\equiv \upsilon=d$, the
Eq.(\ref{ress}) implies that the effective resistances in the
infinite $d$-dimensional lattice $A_1\times...\times A_1$ are
obtained as
$$R_{\alpha\beta^{(l_1...l_d)}}=\frac{2}{\kappa_{\mathbf{l}}}\frac{1}{(2\pi)^d}\int_{0}^{2\pi}dx_1...\int_{0}^{2\pi}dx_d=
\{\kappa_l-2[\cos(l_1x_1+...+l_dx_d)+\small{\mathrm{perm}.}+$$
$$\cos(-l_1x_1+l_2x_2...+l_dx_d)+\small{\mathrm{perm}}.+...+\cos(l_1x_1+...+l_{d-1}x_{d-1}-l_di_d)+\small{\mathrm{perm}.}+$$
$$\cos(-l_1x_1-l_2x_2+l_3x_3+...+l_dx_d)+\small{\mathrm{perm}.}+...+\cos(-l_1x_1-l_2x_2-...-l_{n}x_{n}+l_{n+1}x_{n+1}+...+l_dx_d)+\small{\mathrm{perm}.}]\}$$
\begin{equation}\label{ressdinf}
\{d-2(\cos x_1+\cos x_2+...+\cos x_d)\}^{-1}.
\end{equation}

In the following we consider the special cases of two-dimensional
($d=2$) periodic networks such that in the limit of the large size
of the networks, they tend to the root lattices $A_1\times A_1$ and
$A_2$, respectively. The first case is called finite square network
and the latter one is called finite hexagonal network (although, for
$n> 3$, the underlying networks can be constructed similarly, but
the networks do not possess so physical importance).\\
\textbf{b) Finite square network}\\ For this case ($d=2$), the point
group is the same as the Heisenberg group $H_2\cong (Z_2\times
Z_2)\rtimes Z_2$. More clearly, we have $Z_2\times Z_2=\{e;
S_1\rightarrow S^{-1}_1,S_2\rightarrow S_2; S_1\rightarrow
S_1,S_2\rightarrow S^{-1}_2;S_1\rightarrow S^{-1}_1,S_2\rightarrow
S^{-1}_2\}$ and the third cyclic group $Z_2$ is generated by the
permutation $S_1\leftrightarrow S_2$.

Now, we choose the ordering of elements of $Z_m \times Z_m$ as
follows
\begin{equation}\label{ordering}
V=\{e,a,...,a^{m-1},b,ab,...,a^{m-1}b,...,b^{m-1},ab^{m-1},...,a^{m-1}b^{m-1}\},
\end{equation}
where $a^m=b^m=e$. We use the notation $(k,l)$ for the element
$a^kb^l$ of the group. Clearly, $(k,l)(k',l')=(k+k',l+l')$ and
$(k,l)^{-1}=(-k,-l)$. Then the vertex set $V$ of the network will be
$\{(k,l) : k,l\in\{0,1,...,m-1\}\}$. Then, the corresponding orbits
are given by
\begin{equation}\label{orbit}
P_{k_1k_2}:=O((k_1,k_2)),
\end{equation}
where, $P_{00}=\{(0,0)\}$ (in this case, the partition $P$ is called
homogeneous). In the regular representation of the group, for the
corresponding adjacency matrices and the corresponding idempotents,
we have
$$A_{_{\mathbf{k}=(k_1,k_2)}}=\sum_{g\in
O((k_1,k_2))}g=
S^{k_1}_1S^{k_2}_2+S^{-k_1}_1S^{-k_2}_2+S^{k_2}_1S^{k_1}_2+S^{-k_2}_1S^{-k_1}_2+S^{k_2}_1S^{-k_1}_2+S^{-k_2}_1S^{k_1}_2+$$
\begin{equation}\label{adjacencysq}
S^{-k_1}_1S^{k_2}_2 +S^{k_1}_1S^{-k_2}_2,\;\ \mathrm{for} \;\
k_1\neq k_2,
\end{equation}
$$E_{_{\mathbf{k}=(k_1,k_2)}}=
E_{k_1}\otimes E_{k_2}+E_{-k_1}\otimes E_{-k_2}+E_{k_2}\otimes
E_{k_1}+E_{-k_2}\otimes E_{-k_1}+E_{k_2}\otimes
E_{-k_1}+E_{-k_2}\otimes E_{k_1}+$$
\begin{equation}\label{idemsq}
E_{-k_1}\otimes E_{k_2}+E_{k_1}\otimes E_{-k_2},\;\ \mathrm{for} \;\
k_1\neq k_2
\end{equation}
respectively, and
\begin{equation}\label{adjacencysq'}
A_{_{\mathbf{k}=(k_1,k_1)}}=\sum_{g\in O((k_1,k_1))}g=
S^{k_1}_1S^{k_1}_2+S^{-k_1}_1S^{-k_1}_2+S^{k_1}_1S^{-k_1}_2+S^{-k_1}_1S^{k_1}_2,
\end{equation}
\begin{equation}\label{idemsq'}
E_{_{\mathbf{k}=(k_1,k_1)}}= E_{k_1}\otimes E_{k_1}+E_{-k_1}\otimes
E_{-k_1}+E_{k_1}\otimes E_{-k_1}+E_{-k_1}\otimes E_{k_1}.
\end{equation}
Therefore, the cardinalities of the associate classes $\Gamma_i(o)$
($\kappa_i$), the valencies of the adjacency matrices and the ranks
of the idempotents are given by
$$
\kappa_{_{\mathbf{0}=(0,0)}}=m_{_{\mathbf{0}=(0,0)}}=1,\;\
\kappa_{_{\mathbf{k}=(k_1,k_2)}}=m_{_{\mathbf{k}=(k_1,k_2)}}=8 \;\
\mathrm{for} \;\ 0\neq k_1\neq k_2\neq0,$$
\begin{equation}\label{validemsq}
\mathrm{and}\;\
\kappa_{_{\mathbf{k}=(k,k)}}=m_{_{\mathbf{k}=(k,k)}}=4\;\ ,\;\
\kappa_{_{\mathbf{k}=(k,0)}}=m_{_{\mathbf{k}=(k,0)}}=4.
\end{equation}

The eigenvalues of the adjacency matrix $A_{\mathbf{k}=(k_1,k_2)}$
with $k_1\neq k_2$ are given by
\begin{equation}\label{eigsq}
\lambda^{\mathbf{k}=(k_1,k_2)}_{ij}=2\{\cos
\frac{2\pi(ik_1+jk_2)}{m}+\cos \frac{2\pi(ik_2+jk_1)}{m}+\cos
\frac{2\pi(ik_2-jk_1)}{m}+\cos \frac{2\pi(ik_1-jk_2)}{m}\},
\end{equation}
where for $k_1=k_2$, we have
\begin{equation}\label{eigsq'}
\lambda^{\mathbf{k}=(k_1,k_1)}_{ij}=2\{\cos
\frac{2\pi(i+j)k_1}{m}+\cos \frac{2\pi(i-j)k_1}{m}\}.
\end{equation}
Clearly, for finite square lattice we have $c_{(1,0)}\equiv c=1$ and
$c_{(i_1,i_2)}=0$ for all $i_1\neq 1$ and $i_2\neq0$. Then, by
substituting (\ref{validemsq}) and (\ref{eigsq}) in (\ref{ress}),
the effective resistances on the finite square lattice are given by
\small{{
\begin{equation}\label{resssq}
\hspace{-1cm}R_{\alpha\beta^{(l_1l_2)}}=\frac{1}{m^2\kappa_{_{(l_1l_2)}}}\sum_{k_1,k_2}
\frac{m_{_{(k_1k_2)}}[\kappa_{_{(l_1l_2)}}-2(\cos\frac{2\pi(k_1l_1+k_2l_2)}{m}+\cos\frac{2\pi(k_1l_2+k_2l_1)}{m}+\cos\frac{2\pi(k_1l_2-k_2l_1)}{m}+\cos\frac{2\pi(k_1l_1-k_2l_2)}{m})]}{2-\cos
2\pi k_1/m-\cos 2\pi k_2/m}.
\end{equation}}}
where, $R_{\alpha\beta^{(l_1l_2)}}$ denotes the effective
resistances between $\alpha$ and all the nodes $\beta\in
\Gamma_{_{\mathbf{l}=(l_1l_2)}}(\alpha)$. For instance for $\beta\in
\Gamma_{_{\mathbf{1}=(10)}}(\alpha)$, we obtain
\begin{equation}\label{resssq10}
R_{\alpha\beta^{(10)}}=\frac{1}{2m^2}\sum_{k_1,k_2}
\frac{m_{_{(k_1k_2)}}[2-\cos(2\pi k_1)/m-\cos(2\pi
k_2)/m]}{2-\cos(2\pi k_1)/m-\cos(2\pi
k_2)/m}=\frac{R}{2m^2}\sum_{k_1,k_2}m_{_{(k_1k_2)}}=\frac{1}{2m^2}.m^2=\frac{R}{2}.
\end{equation}

In the limit of the large size of the finite lattice, i.e., in the
limit of $m\rightarrow \infty$, we have the infinite square lattice.
In this limit the eigenvalues (\ref{eigsq}) tend to
$$\lambda^{\mathbf{l}=(l_1l_2)}_{x_1,x_2}=2[\cos(l_1x_1+l_2x_2)+\cos(l_1x_2+l_2x_1)+\cos(l_1x_2-l_2x_1)+\cos(l_1x_1-l_2x_2)]$$
where, $x_1=\lim_{{k_1,m}\rightarrow \infty}2\pi k_1/m$ and
$x_2=\lim_{{k_2,m}\rightarrow \infty}2\pi k_2/m$. Then, the
effective resistances are calculated as follows \small{
\begin{equation}\label{resssqinf}
\hspace{-1cm}R_{\alpha\beta^{(l_1l_2)}}=\frac{1}{8\pi^2}\int_{0}^{2\pi}\int_{0}^{2\pi}
\frac{2-\cos(l_1x_1+l_2x_2)-\cos(l_1x_2+l_2x_1)-\cos(l_1x_2-l_2x_1)-\cos(l_1x_1-l_2x_2)}{2-\cos
x_1-\cos x_2}dx_1dx_2.
\end{equation}}
\textbf{c) Hexagonal network}\\ Now, we consider the finite root
lattice $A_2$ which is called hexagonal lattice. The point group of
the lattice $A_2$ is $S_3\rtimes Z_2$, where $S_3$ is the group of
permutations of the simple roots together with the lowest root (all
permutations of $S_1$, $S_2$ and $(S_1S_2)^{-1}$). With the same
ordering of elements as before, the corresponding orbits are given
by
\begin{equation}\label{orbit}
P_{k_1k_2}:=O((k_1,-k_2)),
\end{equation}
where, $P_{00}=\{(0,0)\}$. Then, for the corresponding adjacency
matrices and the corresponding idempotents, we have
$$A_{_{\mathbf{k}=(k_1,k_2)}}=\sum_{g\in
O((k_1,-k_2))}g=
S^{k_1}_1S^{-k_2}_2+S^{-k_1}_1S^{k_2}_2+S^{-k_2}_1S^{k_1}_2+S^{k_2}_1S^{-k_1}_2+S^{-k_2}_1S^{k_1+k_2}_2+S^{k_2}_1S^{-(k_1+k_2)}_2+$$
\begin{equation}\label{adjacencyhex}
S^{k_1+k_2}_1S^{k_2}_2+S^{-(k_1+k_2)}_1S^{-k_2}_2+
S^{k_2}_1S^{k_1+k_2}_2+S^{-k_2}_1S^{-(k_1+k_2)}_2+
S^{k_1+k_2}_1S^{k_1}_2+ S^{-(k_1+k_2)}_1S^{-k_1}_2,\;\ \mathrm{for}
\;\ k_1\neq k_2,
\end{equation}
$$E_{_{\mathbf{k}=(k_1,k_2)}}=
E_{k_1}\otimes E_{-k_2}+E_{-k_1}\otimes E_{k_2}+E_{-k_2}\otimes
E_{k_1}+E_{k_2}\otimes E_{-k_1}+E_{-k_2}\otimes
E_{k_1+k_2}+E_{k_2}\otimes E_{-(k_1+k_2)}+$$
\begin{equation}\label{idemhex}
E_{k_1+k_2}\otimes E_{k_2}+E_{-(k_1+k_2)}\otimes
E_{-k_2}+E_{k_2}\otimes E_{k_1+k_2}+E_{-k_2}\otimes
E_{-(k_1+k_2)}+E_{k_1+k_2}\otimes E_{k_1}+E_{-(k_1+k_2)}\otimes
E_{-k_1},\;\ \mathrm{for} \;\ k_1\neq k_2
\end{equation}
respectively, and
$$
A_{_{\mathbf{k}=(k,k)}}=\sum_{g\in O((k,k))}g=
S^{k}_1S^{-k}_2+S^{-k}_1S^{k}_2+S^{k}_1S^{-2k}_2+S^{-k}_1S^{2k}_2+S^{2k}_1S^{-k}_2+S^{-2k}_1S^{k}_2+$$
\begin{equation}\label{adjacencyhex'}
S^{2k}_1S^{k}_2+S^{-2k}_1S^{-k}_2+S^{k}_1S^{2k}_2+S^{-k}_1S^{-2k}_2,
\end{equation}
$$
E_{_{\mathbf{k}=(k,k)}}= E_{k}\otimes E_{-k}+E_{-k}\otimes
E_{k}+E_{k}\otimes E_{-2k}+E_{-k}\otimes E_{2k}+E_{2k}\otimes
E_{-k}+E_{-2k}\otimes E_{k}+$$
\begin{equation}\label{idemhex'}
E_{2k}\otimes E_{k}+E_{-2k}\otimes E_{-k}+E_{k}\otimes
E_{2k}+E_{-k}\otimes E_{-2k}.
\end{equation}
Therefore, the cardinalities of the associate classes $\Gamma_i(o)$
($a_i$), the valencies of the adjacency matrices and the ranks of
the idempotents are given by
$$
\kappa_{_{\mathbf{0}=(0,0)}}=m_{_{\mathbf{0}=(0,0)}}=1,\;\
\kappa_{_{\mathbf{k}=(k_1,k_2)}}=m_{_{\mathbf{k}=(k_1,k_2)}}=12 \;\
\mathrm{for} \;\ 0\neq k_1\neq k_2\neq0,$$
\begin{equation}\label{validemhex}
\mathrm{and}\;\
\kappa_{_{\mathbf{k}=(k,k)}}=m_{_{\mathbf{k}=(k,k)}}=10\;\ ,\;\
\kappa_{_{\mathbf{k}=(k,0)}}=m_{_{\mathbf{k}=(k,0)}}=6.
\end{equation}

The eigenvalues of the adjacency matrix $A_{\mathbf{k}=(k_1,k_2)}$
with $k_1\neq k_2$ are given by
$$
\lambda^{(\mathbf{k})}_{ij}=2\{\cos \frac{2\pi(ik_1-jk_2)}{m}+\cos
\frac{2\pi(ik_2-jk_1)}{m}+\cos \frac{2\pi(ik_2-j(k_1+k_2))}{m}+\cos
\frac{2\pi(i(k_1+k_2)-jk_2)}{m}+$$
\begin{equation}\label{eighex}
\cos \frac{2\pi(ik_2+j(k_1+k_2))}{m}+\cos
\frac{2\pi(i(k_1+k_2)+jk_1)}{m}\}.
\end{equation}
where for $k_1=k_2\equiv k$, we have
\begin{equation}\label{eigsq'}
\lambda^{\mathbf{k}=(k,k)}_{ij}=2\{\cos \frac{2\pi(i-j)k}{m}+\cos
\frac{2\pi(i-2j)k}{m}+\cos \frac{2\pi(2i-j)k}{m}+\cos
\frac{2\pi(2i+j)k}{m}+\cos \frac{2\pi(i+2j)k}{m}\}.
\end{equation}
Then, similar to the case of finite square lattice, one can
calculate effective resistances $R_{\alpha\beta^{(l)}}$.

In the limit of $m\rightarrow \infty$, we have the infinite
hexagonal lattice. In this limit the eigenvalues (\ref{eigsq}) tend
to
$$\lambda^{\mathbf{l}=(l_1l_2)}_{x_1,x_2}=2[\cos(l_1x_1-l_2x_2)+\cos(l_1x_2-l_2x_1)+\cos(l_1x_2-l_2(x_1+x_2))+\cos(l_1(x_1+x_2)-l_2x_2)+$$
$$\cos(l_1x_2+l_2(x_1+x_2))+\cos(l_1(x_1+x_2)+l_2x_1)]$$ where,
$x_1$ and $x_2$ are defined as before. Then, the effective
resistances are calculated as follows
$$R_{\alpha\beta^{(l_1l_2)}}=\frac{1}{8\pi^2}\int_{0}^{2\pi}\int_{0}^{2\pi}
\frac{3-\cos(l_1x_1-l_2x_2)-\cos(l_1x_2-l_2x_1)-\cos(l_1x_2-l_2(x_1+x_2))-
}{3-\cos x_1-\cos x_2-\cos(x_1+x_2)}$$
\begin{equation}\label{resssqinf}
\hspace{-1cm}
\frac{\cos(l_1(x_1+x_2)-l_2x_2)-\cos(l_1x_2+l_2(x_1+x_2))-\cos(l_1(x_1+x_2)+l_2x_1)}{3-\cos
x_1-\cos x_2-\cos(x_1+x_2)}dx_1dx_2.
\end{equation}
\subsubsection{An example of the underlying  networks of group
association schemes} In this subsection we consider
underlying network of group association scheme $S_n$ and .\\
\textbf{Symmetric group $S_n$}\\ The symmetric group $S_n$ is
ambivalent in the sense that its conjugacy classes are real, i.e.,
$C_i={C_i}^{-1}$ for all $i$ and so form a symmetric association
scheme.

As it is well known, for the group $S_n$, conjugacy classes are
determined by the cycle structures of elements when they are
expressed in the usual cycle notation. The useful notation for
describing the cycle structure is the cycle type
$[\nu_1,\nu_2,...,\nu_n]$ , which is the listing of number of cycles
of each length (i.e, $\nu_1$ is the number of one cycles, $\nu_2$ is
that of two cycles and so on).  Thus, the number of elements in  a
conjugacy class or stratum is given by
\begin{equation}
|C_{[\nu_1,\nu_2,...,\nu_n]}|=\frac{n!}{\nu_1!2^{\nu_2}\nu_2!...n^{\nu_n}\nu_n!}.
\end{equation}
On the other hand a partition $\lambda$ of $n$ is a sequence
$(\lambda_1, . . . , \lambda_n)$ where $\lambda_1\geq\cdots\geq
\lambda_n$ and $\lambda_1 +\cdots+\lambda_n = n$, where in terms of
cycle types
\begin{equation}
\lambda_1=\nu_1+\nu_2+\cdots+\nu_n, \;\;\;\
\lambda_2=\nu_2+\nu_3+\cdots+\nu_n, \;\;\;\
 \cdots, \;\;\;\
\lambda_n=\nu_n.
\end{equation}
 The notation $\lambda\vdash n$  indicates that $\lambda$ is a
partition of $n$. There is one conjugacy class for each partition
$\lambda\vdash n$ in $S_n$, which consists of those permutations
having cycle structure described by $\lambda$. We denote by
$C_\lambda$ the conjugacy class of $S_n$ consisting of all
permutations having cycle structure described by $\lambda$.
Therefore the number of conjugacy classes of $S_n$, namely diameter
of its scheme is equal to the number of partitions of $n$, which
grows approximately by $ \frac{1}{4\pi\sqrt{3}}e^{\pi\sqrt{2n/3}}$.

We consider the case where the generating set consists of the set of
all transposition, i.e, $C_1=C_{[2,1,1,1,1...,1]}$. For the
characters at the transposition, it is known that \cite{ri}
\begin{equation}\label{eigen2}
\chi_\lambda(\alpha_1)=\frac{2!(n-2)!dim(\rho_\lambda)}{n!}\sum_{j}\left(\left(
\begin{array}{cc}
 \lambda_j \\ 2
 \end{array}
\right)- \left(
\begin{array}{cc}
 \lambda'_j \\ 2
 \end{array}
\right) \right).
\end{equation}
Here, $\lambda'$ is the partition generated by transposing the Young
diagram of $\lambda$, while $\lambda'_j$ and $\lambda_j$ are the
$j$-th components of the partitions $\lambda'$ and $\lambda$, and
$\rho_\lambda$ is the irreducible representation corresponding to
partition $\lambda$.

Then the eigenvalues of the adjacency matrix can be written as
\begin{equation}\label{eigen2}
P_{\lambda
1}=\frac{d_{\lambda}k_1}{m_\lambda}\chi_\lambda(\alpha_1)=\sum_{j}\left(\left(
\begin{array}{cc}
 \lambda_j \\ 2
 \end{array}
\right)- \left(
\begin{array}{cc}
 \lambda'_j \\ 2
 \end{array}
\right) \right).
\end{equation}
In the above calculation, we have used the following results for the
characters of the $n$-cycles
\[
\chi_\lambda \left((n)\right) = \left\{
\begin{array}{ll}
(-1)^{n-k} & \mbox{for $\lambda = (k,1,\ldots,1)$, $k \in \{1,\ldots,n\} $ }\\
\hspace{3mm} 0 & \mbox{otherwise}
\end{array}
\right.
\]
and
\[
\chi_{(k,1,\cdots,1)}\left(\mathrm{id} \right) =
{dim}(\rho_{(k,1,\cdots,1)})=\left(
\begin{array}{cc}
 n-1 \\ k-1
 \end{array}
\right), \;\;\;\ P_{\lambda1}=\frac{1}{2}(2nk-n^2-n).
\]
Then, one can evaluate effective resistances by using the Eq.(5-80).
In the following, we consider the underlying network of group
association scheme $S_4$ with diameter $d=4$, in details. To do so,
we use the conjugacy classes of $S_4$ given by Eq.(\ref{cons4}) and
the adjacency matrices $A_i=\bar{C_i}, \;\ i=0,1,...,4$ which
satisfy the following Bose-Mesner algebra
$$
A^2=6A_0+3A_2+2A_3, \;\;\ AA_2=4A+4A_4, \;\;\ AA_3=A+2A_4, \;\;\
AA_4=4A_2+4A_3,
$$
$$
A_2^2=8A_0+4A_2+8A_3, \;\;\ A_2A_3=3A_2, \;\;\ A_2A_4=4A+4A_4,\;\;\
A_3^2=3A_0+2A_3,
$$
\begin{equation}\label{s4}
A_3A_4=2A+A_4, \;\;\ A_4^2=6A_0+3A_2.
\end{equation}
By using the character table of the group $S_4$ and Eq.
(\ref{group}), one can obtain
$$P_{0k}=1,\;\ k=0,...,4,\;\;\
P_{10}=P_{12}=P_{13}=-P_{11}=-P_{14}=6,\;\;\ P_{20}=P_{23}=8,\;\;\
P_{21}=P_{24}=0,$$
\begin{equation}\label{s4p}
P_{22}=-4,\;\ P_{30}=3,\;\ P_{31}=-P_{33}=-P_{34}=1,\;\ P_{32}=0,\;\
P_{40}=6,\;\ P_{41}=P_{43}=-P_{44}=-2,\;\ P_{42}=0.
\end{equation}
Now, by using the Eq.(\ref{ress}), we obtain
$$\hspace{-5.5cm}R_{\alpha\beta^{(1)}}=\frac{1}{6}\{\frac{1}{12c_1+2c_3+8c_4}+\frac{9}{12c_1+8c_2+4c_3+4c_4}\},$$
$$R_{\alpha\beta^{(2)}}=\frac{1}{36}\{\frac{3}{12c_1+2c_3+8c_4}+\frac{20}{12c_2+3c_3+6c_4}-\frac{9}{4c_3+8c_4}+\frac{27}{12c_1+8c_2+4c_3+4c_4}\},$$
$$\;\ R_{\alpha\beta^{(3)}}=\frac{1}{18}\{\frac{1}{12c_1+2c_3+8c_4}+\frac{6}{12c_2+3c_3+6c_4}+\frac{18}{4c_3+8c_4}+\frac{18}{12c_1+8c_2+4c_3+4c_4}+\},$$
\begin{equation}
\hspace{0.15cm}R_{\alpha\beta^{(4)}}=\frac{1}{48}\{\frac{5}{12c_1+2c_3+8c_4}+\frac{16}{12c_2+3c_3+6c_4}+\frac{45}{4c_3+8c_4}+\frac{27}{12c_1+8c_2+4c_3+4c_4}\},
\end{equation}
where, $R_{\alpha\beta^{(i)}}$ denotes the effective resistance
between the node $\alpha$ and all nodes $\beta\in
\Gamma_{i}(\alpha)$ for $i=1,2,3,4$.
\section{Conclusion}
Based on stratification of underlying networks of association
schemes and using their algebraic combinatoric structure such as
Bose-Mesner algebra together with spectral techniques, evaluation of
effective resistances on these networks was discussed. It was shown
that, in these types of networks, the effective resistances between
a node $\alpha$ and all nodes $\beta$ belonging to the same stratum
with respect to $\alpha$ are the same. Then, by assumption that all
of the conductances except for one of them is zero, a procedure for
evaluation of effective resistances on particular underlying
networks for which all of adjacency matrices are written as
polynomials of the first adjacency matrix $A$ of the network, was
given such that effective resistances can be evaluated without using
the spectrum of the networks. Moreover, an explicit analytical
formula for effective resistance between arbitrary nodes
$\alpha,\beta$ of an underlying resistor network of an association
schemes (where all of conductances are non-zero) was given in terms
of the spectrum of the networks. In each case, evaluation of
effective resistance on some important finite underlying networks of
association schemes and their corresponding infinite networks was
given.
\newpage
 \vspace{1cm}\setcounter{section}{0}
 \setcounter{equation}{0}
 \renewcommand{\theequation}{A-\roman{equation}}
  {\Large{Appendix A}}\\
In this appendix we show that, for underlying networks of
association schemes with diameter $d$ such that the adjacency matrix
$A$ (or any of the other adjacency matrices $A_i$, $i=2,...,d$ which
gives a connected network) has $d+1$ distinct eigenvalues, all of
the adjacency matrices are polynomials of $A$, i.e., $A_i=P_i(A)$,
where $P_i$ is not necessarily of degree $i$. To do so, let $A$ be
the adjacency matrix of the connected network with $d+1$ distinct
eigenvalues $P_{1k}$, $k=0,1,...,d$. Then, by using the structure of
the Bose-Mesner algebra, i.e., Eq.(\ref{m2}), one can write
$$A^l=\sum_{k=0}^{d}(P_{1k})^lE_k,$$
or in the matrix form
$$(\begin{array}{ccccc}
               \mathbf{1} & A & A^2 & \ldots & A^d
             \end{array}
  )^t=V(\begin{array}{ccccc}
               E_0 & E_1 & E_2 & \ldots & E_d
             \end{array})^t
$$
where, $V$ is the Vandermonde matrix  $$V=\left(\begin{array}{cccc}
                                            1 & 1 & \ldots & 1 \\
                                            P_{10} & P_{11} & \ldots & P_{1d} \\
                                            P^2_{10} & P^2_{11} & \ldots & P^2_{1d} \\
                                            \vdots & \vdots & \vdots & \vdots \\
                                            P^d_{10} & P^d_{11} & \ldots & P^d_{1d}
                                          \end{array}\right).
$$
Clearly $V$ is invertible due to the distinctness of the eigenvalues
$P_{1k}$ for $k=0,1,...,d$. Then, we have
$$(\begin{array}{ccccc}
               E_0 & E_1 & E_2 & \ldots & E_d
             \end{array}
  )^t=V^{-1}(\begin{array}{ccccc}
              \mathbf{1} & A & A^2 & \ldots & A^d
             \end{array})^t.
$$
Now, by using (\ref{m2}), we write the idempotents $E_i$ in terms of
$A_i$ to obtain
$$(\begin{array}{ccccc}
               E_0 & E_1 & E_2 & \ldots & E_d
             \end{array}
  )^t=\frac{1}{n}Q(\begin{array}{ccccc}
              \mathbf{1} & A & A_2 & \ldots & A_d
             \end{array})^t=V^{-1}(\begin{array}{ccccc}
              \mathbf{1} & A & A^2 & \ldots & A^d
             \end{array})^t.
$$
Therefore, the adjacency matrices $A_i$, $i=0,1,...,d$ can be
written as polynomials of $A$, i.e., we have
$$(\begin{array}{ccccc}
              \mathbf{1} & A & A_2 & \ldots & A_d
             \end{array})^t=n(VQ)^{-1}(\begin{array}{ccccc}
              \mathbf{1} & A & A^2 & \ldots & A^d
             \end{array})^t.
$$

\end{document}